\documentclass[12pt,preprint]{aastex}

\slugcomment{To appear on AJ}

\def\gsim{\;\lower.6ex\hbox{$\sim$}\kern-7.75pt\raise.65ex\hbox{$>$}\;}
\def\lsim{\;\lower.6ex\hbox{$\sim$}\kern-7.75pt\raise.65ex\hbox{$<$}\;}

\shorttitle{Abundances in the globular cluster NGC 2808}
\shortauthors{E. Carretta}

\begin{document}


\title{Abundances in red giant stars of NGC 2808 and correlations between
chemical anomalies and global parameters in globular clusters\altaffilmark{1}}


\author{Eugenio Carretta}
\affil{INAF-Osservatorio Astronomico di Bologna, via Ranzani 1, I-40127
 Bologna, ITALY}
\email{eugenio.carretta@bo.astro.it}


\altaffiltext{1}{Based on data collected at the European Southern Observatory,
Chile, during the FLAMES Science Verification program with the UVES
spectrograph at VLT-UT2.}

\begin{abstract}
We present the abundance analysis of stars from the tip of the red giant branch
(RGB) to below the RGB-bump in the globular cluster NGC 2808, based on high 
resolution echelle spectra.
We derived abundances of Al, $\alpha$-process elements (Si I, Ca I, Ti I  and
Ti II) and Fe-group elements (Sc II, V I, Cr I, Cr II, Mn I, Co I, Ni I). Apart
from Mg being somewhat reduced, likely because it has
been depleted at the  expense of
Al in the MgAl cycle, the other $\alpha$-element ratios show the overabundance 
typical of halo stars of similar metallicity. Mn is underabundant, whereas 
Fe-group elements have typical abundance ratios near the solar value. We detect
star-to-star differences in Al abundances from the RGB tip down to the faintest
star below the RGB-bump, correlated with Na abundances at all luminosities. The
slope of the Na-Al correlation is similar to the one found in M~13 by Sneden et
al. (2004), but it is different from those in other globular clusters of
similar metallicity. We find
that the amount of chemical inhomogeneities along the Na-O and Mg-Al
anticorrelations in globular cluster red giants is correlated with the present
day cluster mass and ellipticity. Moreover, we find for the first time a 
correlation between the spread in proton-capture elements and orbital
parameters of clusters. The chemical anomalies are more extended in clusters
having large-sized orbits and longer periods, and in clusters with larger
inclination angles of the orbit with respect to the Galactic plane.

\end{abstract}


\section{Introduction}

Galactic globular clusters (GCs) are a real gold mine to dig out important
informations in a variety of astrophysical issues ranging from dating the
Universe through stellar clocks, to stellar evolution studies, to the formation
and early chemical evolution of the Galaxy. Nevertheless, little is known
about $their$ origin and formation processes. Typical present-day masses for
GCs are tantalizingly close to those characteristic of protogalactic clouds as
estimated, e.g., from the surprisingly small scatter observed in Mg and other
elements in extremely metal-poor stars \citep{car02}. However, the exact
formation mechanism is still unknown, even if a few tentative models have
been proposed, starting with the pioneering work by \citet{cay86} (see also
\citealt{par04} and references therein).

Fortunately, the detailed chemical composition of their stars can help us with
this issue: in low-mass, long-living GC stars there is a ``black box" where
the conditions existing when the clusters converted its gas into
stars are recorded (apart from a few alterations due to 
stellar evolution).

Although present theoretical models seem to show a lag in  (at least
quantitatively) correctly reproducing all the observations, abundance analyses
are unveiling a more and more defined, though complex, pattern. In particular, 
light elements (C, N, O, Na, Mg, Al) involved in proton-capture synthesis seem to
play a key role in shaping the overall chemical composition of GC stars (see
the review by \citealt{gra04} for a summary and updated references).
The discovery of anticorrelations (and correlations) between elements 
arising from $p-$capture
chains such as NeNa and MgAl in unvevolved cluster stars
\citep{gra01,car0447t,ram02,ram03} provides a fundamental piece of information.
The temperature in the H-burning shell of low mass stars presently observed
in GCs cannot be high enough to efficiently operate the MgAl chain
\citep{lan95}; moreover, unevolved stars burn H only in their cores and do not have
convective envelope able to mix out products of inner nuclear processes.
Thus the inference is that this synthesis took place in a prior generation of
more massive stars already evolved and died in the few 10$^8$ yrs since
the cluster formation (see \citealt{car05} for a detailed discussion about the
mass ranges of likely candidates).

Hence, these light elements are direct witnesses of the very early
phases in the lifetime of a GC and deriving accurate
abundances of $p-$capture elements in a large number of stars for an extensive
sample of GCs is a crucial goal.

The target of the present study, NGC 2808, is a peculiar cluster of
intermediate metallicity, 
famous for the bimodal distribution of stars on the horizontal branch (HB), 
maybe related to the He content imprinted early in the cluster life (see for
instance \citealt{cal04}). Despite its peculiarity, NGC 2808 had not received a
great deal of attention in stellar abundance analyses until very recently, but
is now becoming one of the most well studied clusters. \citet{car03} analyzed
more than 80 red giants to derive Na abundances, unveiling for the first time
in this cluster large star-to-star variations in the abundance of a $p-$capture
element. Afterwards, \citet{car04nao28} found that the well-known Na-O
anticorrelation discovered by the Lick-Texas group (exhaustive references in
\citealt{gra04}) is also present in NGC 2808, where large levels of O depletion
and Na enhancement are observed. The study of the  Na-O signature will be soon
extended to a sample of 130 stars in a forthcoming paper (Carretta et al. 2005,
in preparation).

Nevertheless, nothing has been said yet about the elements heavier than Na. In the
present work we fill this gap by studying the detailed composition of 19 RGB
stars and by deriving abundances for species from Al to Ni. A comparison with 
M 5 and M 4, clusters with metallicity
very similar to the value [Fe/H]$=-1.14$ dex\footnote{We adopt the usual 
spectroscopic notation, $i.e.$ 
[X]= log(X)$_{\rm star} -$ log(X)$_\odot$ for any abundance quantity X, and 
log $\epsilon$(X) = log (N$_{\rm X}$/N$_{\rm H}$) + 12.0 for absolute number
density abundances.} found for NGC 2808 \citep{car04nao28}, will greatly help to 
highlight similarities and differences in the abundance patterns, since analyses
of comparable resolution, quality and sample size are available for these three
clusters.

Finally, we put together a sample of well studied GCs, with abundance
analyses from high resolution spectra of at least 20 stars per cluster, to
search for correlations between chemical inhomogeneities and global cluster
parameters. We show that the amount of spread along the Na-O and Mg-Al
anticorrelations seems to be somewhat related to the present day masses and
ellipticities of GCs. Moreover, we uncovered for the first time a well defined
relation between the chemical anomalies in elements originated from $p-$capture reactions
and orbital parameters of the clusters.

The paper is organized as follows: Sections 2, 3 and 4 are devoted to the
observations and analysis; Section 5 presents the derived abundances and in
Section 6 we define a quantitative estimate of the amount of chemical 
inhomogeneity in a cluster and its possible correlation with chemistry,
structural and orbital cluster parameters.

\section{Observations and data reduction}

The program stars were observed during the FLAMES Science Verification program
at the ESO Paranal Observatory in 2003, January 24-25.
Thirteen RGB stars were selected from the database of \citet{bed00} in the 
uppermost 0.8 mag interval from the tip ($V=13.2$). The other 7
targets were chosen to sample the 2.5 fainter magnitudes; the two faintest
stars are located just below the RGB-bump (at $V \sim 16.15$ in this cluster,
\citealt{bed00}).
All these stars were observed with the fiber fed Red Arm of the high resolution
spectrograph UVES ($R \sim 47000$, fibers of 1 arcsec entrance aperture,
grating centered at 5800~\AA\ with spectral coverage of about 2000~\AA). 

NGC 2808 is quite concentrated and our targets are within a 7$\arcmin$ radius
from the cluster center; all stars were chosen to be free from companions
closer than 2.4 arcsec and brighter than $V+1.5$, where $V$ is the target
magnitude.

Data reduction of the UVES spectra, using an ad hoc Data Reduction Software 
\citep{mul02}, included bias subtraction, flat-fielding correction, correction
for scattered light,  spectra extraction and wavelength calibration using a
reference Th-Ar  calibration lamp. The reduction is described at length in
\citet{cac04}, where details of the observations can be found, as well as
magnitudes and coordinates of the program stars.

Multiple exposures for the same stars were coadded, after a shift to zero
radial velocity, to enhance the $S/N$ values.
However, the $S/N$ ratios vary a lot, since the selection of stars in
the Science Verification program was
optimized for studying mass loss, not for abundance analysis. 
Final $S/N$ values are listed in Table~1 of \citet{car04nao28}. As found in that
study, the spectrum of star 34013 (below the magnitude of the RGB-bump, with 
a $S/N \sim20$) was mostly useless to derive reliable equivalent
widths, so this star was dropped from further analysis.

\section{Atmospheric parameters and Iron abundances}

The procedure adopted to derive effective temperatures and surface gravities
is reported in detail in \citet{cac04}. Briefly summarizing, 
we used $K$ magnitudes taken from the Point
Source Catalogue of 2MASS \citep{cut03} and
transformed to the TCS photometric system.

We obtained T$_{\rm eff}$'s and bolometric corrections 
from dereddened $V-K$ colors, by employing
the relations by \citet{alo99}. 
Surface gravities log $g$'s were obtained from effective temperatures and 
bolometric corrections, assuming that the stars have masses of 0.85
M$_\odot$. The adopted  bolometric magnitude of the Sun is 
$M_{\rm bol,\odot} =4.75$. 

We adopted a distance modulus of $(m-M)_V$=15.59
and a reddening of $E(B-V)$ = 0.22 (from \citealt{har96})\footnote{As updated 
at {\tt http://physun.physics.mcmaster.ca/Globular.html}}, and the relations 
$E(V-K) = 2.75 E(B-V)$, $A_V = 3.1 E(B-V)$, and $A_K = 0.353 E(B-V)$ 
\citep{card89}. The magnitudes, colors and derived parameters for all stars in
our sample are listed in \citet{cac04}.

In \citet{cac04} an average internal error of 70~K in temperature was
derived over the magnitude range sampled by the FLAMES Science Verification
program. This is likely an overestimate of the actual error 
affecting the
brighter stars in the present study: judging from photometric errors quoted by 
\citet{bed00} and those associated to 2MASS $K$ 
magnitudes\footnote{See  
www.ipac.caltech.edu/2mass/releases/allsky/doc/sec2\_2.html\#pscphotprop},
a more realistic estimate would be $\sim 44$~K for this group of stars
(including a contribution from errors in reddening),
increasing up to 52 K for the faintest stars in our sample, where 2MASS errors
increase from 0.02 mag to 0.04 mag. However, because the observed
stars cover such a large range in magnitude, in the following we keep 70~K as a
conservative estimate of random errors in T$_{\rm eff}$. 

\begin{figure}
\epsscale{.90}
\plotone{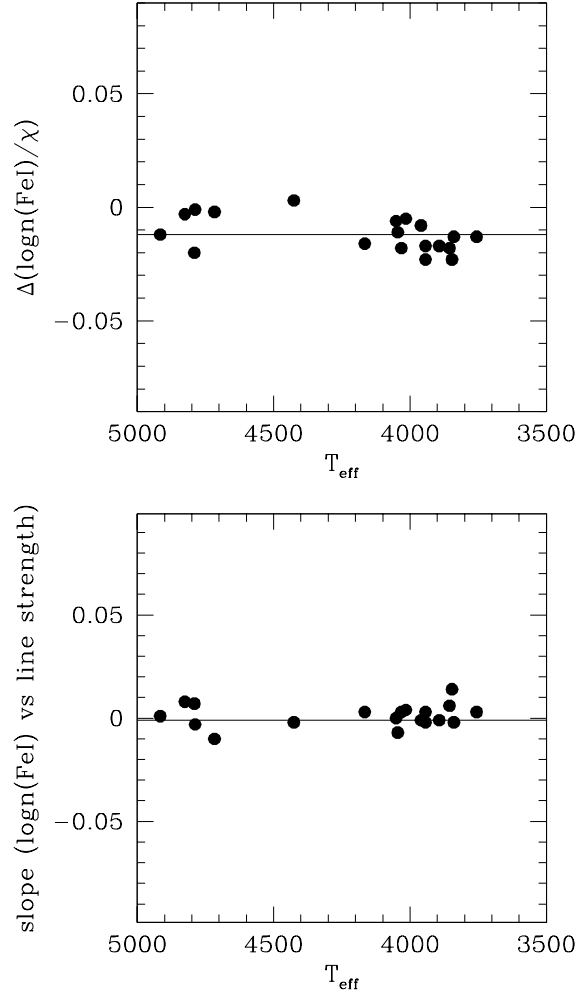}
\caption{Upper panel: slopes of the relationship resulting between the Fe I
abundances and the excitation potential $\chi$ for individual stars in the
sample as a function of the adopted T$_{\rm eff}$'s from photometry.
Lower panel: slopes in the relation between expected line strength and
abundances of Fe I for individual stars, as a function of the adopted 
T$_{\rm eff}$'s. In both panels, solid lines represent the average values for
the sample.\label{f01}}
\end{figure}

In the upper panel of Figure~\ref{f01} we plot the values of the slopes of
the abundances from neutral Fe I lines with respect to the excitation potential
$\chi$ for individual stars in our sample, as a function of the adopted 
T$_{\rm eff}$'s from photometry. This Figure supports the reliability of 
the adopted temperature scale: it seems to reproduce quite well the 
excitation equilibrium within the
errors: the average value of the slope 
$\Delta(\log n(FeI)/\chi) = -0.012 \pm 0.002$ dex/eV 
with $\sigma=0.008$ dex/eV corresponds to about 60 K. 

Since surface gravities are derived from the position of stars in the
color-magnitude diagram, internal errors should take into account uncertainties
in bolometric corrections, distance moduli, adopted temperatures and
masses. Summing in quadrature all these contributions 
(see e.g. \citealt{car0447t}) we find internal errors not larger than 0.1 dex 
associated to the adopted gravities. The good ionization equilibrium found
for Fe \citep{car04nao28} supports the adopted values of $\log g$. 

Microturbulent velocities $v_t$ were derived for each star by eliminating
trends in the relations between expected line strength (see \citealt{mag84})
and abundances from neutral Fe lines. In the lower panel of
Figure~\ref{f01} we show the resulting slopes for the sample. Random error 
of $\pm 0.10$ km s$^{-1}$ can be estimated from typical uncertainties in the
slopes of the above relationships.

Finally, we interpolated in the \citet{kur93} grid of model atmospheres (with
the overshooting option set on) choosing the model with the appropriate
atmospheric parameters whose metal abundance was the best match to the derived
Fe abundance. Internal errors in metallicity, as evaluated from the star-to-star
scatter, are about 0.05-0.06 dex.

Table~\ref{tab1}, reproduced from \citet{car04nao28}, lists the stellar
atmospheric parameters and the $S/N$ ratios of our sample of stars in 
NGC 2808, as well as the derived abundances of [Fe/H] from Fe I and 
Fe II. 
The 1$\sigma$ rms value about the mean abundance is also listed for each stars,
together with the number $n$ of lines used.

\section{Analysis}

The abundances were derived using the atmospheric parameters discussed in the
previous Section, the \citet{kur93} set of stellar atmospheres and measured
equivalent widths ($EW$s) for features of Al, of $\alpha$-process elements (Mg I,
Si I, Ca I, Ti I and Ti II) and of Fe-group elements (Sc II, V I, Cr I, Cr II,
Mn I, Co I, Ni I, in addition to Fe I and Fe II). Abundances of oxygen from the 
spectrum synthesis of the
forbidden [O I] lines at 6300-63~\AA\ and of Na from $EW$s measured on these
same spectra were presented in \citet{car04nao28}.

The $EW$s\footnote{$EW$s are available only in the electronic edition of the 
Astronomical Journal.} were 
measured on the spectra using automatic routines of the
recently updated ROSA package \citep{gra88}, with Gaussian fits to measured
profiles and iterative clipping to derive a local continuum around each line.
The procedure is explained in details in \citet{bra01}.
The list of lines and their atomic parameters are those given in \citet{gra03};
in particular the updated treatment of collisional damping from \citet{bar00}
is used, when available. Reference solar abundances are as in \citet{gra03}.

The abundance ratios [X/Fe] are summarized in Table~\ref{tab2} and 
Table~\ref{tab3}, and are computed using Fe abundances from Fe I lines
for neutral species and from Fe II lines for ionized species. For each species
the number of measured lines, the average abundance ratio for 
individual stars, and the 1$\sigma$ rms, i.e. the standard deviation about
the mean abundance, are listed.
Our Al abundances rest on the doublet at 6696-98~\AA, the only feature for this
species falling in the covered spectral range and are listed in 
Table~\ref{tab2}.

Among the $\alpha$-elements we derived abundances of Ti from lines in two
different stages of ionization: this allows us an additional check
for evidence of departures from the LTE assumption (we recall that 
\citealt{car04nao28} did not find any convincing evidence in their analysis 
of Fe abundances).
On average, we found [Ti/Fe]II - [Ti/Fe]I $=-0.11 \pm 0.02$, with an rms=0.08
dex (19 stars) and no trend with temperature.  
Moreover, almost half
of this difference might be explained as due to typical internal errors in
temperature (see below) affecting the ionization equilibrium of Ti.
Hence, we do not regard this average difference as very significant.
Abundance ratios of $\alpha$-process elements are given in Table~\ref{tab2}.

Besides Fe, lines of several elements of the Fe-group were measured.
Corrections for elements with not negligible 
hyperfine structure splitting (Sc II, V I, Mn I) were applied; 
references are given in \citet{gra03}. For the ionization equilibrium of Cr
we found on average [Cr/Fe]II - [Cr/Fe]I $=+0.06 \pm 0.02$, with an rms=0.09
dex (18 stars), again with no trend with T$_{\rm eff}$. 
A typical random error of $\pm 70$~K is able by itself to produce
such a shift in the Cr ionization equilibrium. Again, we regard this result
as a good support for the adopted gravities and a fair argument against
the existence of strong departures from the LTE assumption.
Abundances of Fe-group elements are listed in Table~\ref{tab3}.

Table~\ref{tab4} shows the sensitivity of the derived abundance ratios
to variations in the adopted atmospheric parameters for Fe and the other
elements in the present study. The entries in this Table were obtained by
re-iterating the analysis while changing each time only one of the parameters.
The variation in the parameter (shown in the table heading) was chosen to
be equal to the typical random error previously estimated for the parameter;
hence, this table represents the sensitivity of the abundances to the
$actual$ uncertainties in the adopted atmospheric parameters.
Column 7 of Table~\ref{tab4} allows us to estimate the effect of errors
in the $EW$s; this was obtained by weighting the error in the abundance derived
from an individual line (0.116 dex, from the average error for Fe I over all
stars) with the square root of the mean number of lines measured for each
element. The total error bar is given in the last column as the quadratic sum
of all the contributions.
This exercise was done for star 48889 (among the brightest ones,
near the RGB tip) and for star 43217, the faintest in our sample.

\section{Derived abundances}

Mean abundances for individual elements in the stars of NGC 2808 are given in 
Table~\ref{tab5}, where also the average values found by
\citet{car04nao28} for Na, O and Fe are listed; Na abundances include corrections for
departures from the LTE assumption according to \citet{gra99}.
For each average abundance we give the number of stars used in the
average (column 2) and the observed star-to-star scatter (standard deviation about
the mean abundance, column 3). This scatter might be compared with the total
uncertainty as derived from Table~\ref{tab4} and listed in column 4: this
value $\sigma_{exp}$ can be regarded as the predicted total error expected from
the uncertainties in the atmospheric parameters combined with errors in $EW$
measurements.

A quantitative test of the presence of intrinsic star-to-star variations
in a given ratio may be obtained from the comparison of values in 
columns 3 and 4
(this approach is very similar to the so-called spread ratio introduced by
\citealt{cm05}). For most species, the observed star-to-star scatter is equal
or even lower than expected on the basis of the error analysis. Apart from Sc II
and Mn, where the small number of lines and the effects of HFS correction might
combine to enhance the observed scatter, the elements clearly standing out are
O, Na, Al and Mg. According to Table~\ref{tab4} and Table~\ref{tab5}
these are the species whose range of abundances 
among stars on the RGB in NGC 2808 can be regarded as real star-to-star 
variations.
Of course, this is exactly what is expected, because these are the elements 
involved in the $p-$capture reactions in H-burning at high-temperature  (see
\citealt{gra04} for references) whose action is well known to provide a large
spread in these abundance ratios in every globular cluster studied up to date.

\begin{figure}
\epsscale{1.0}
\plotone{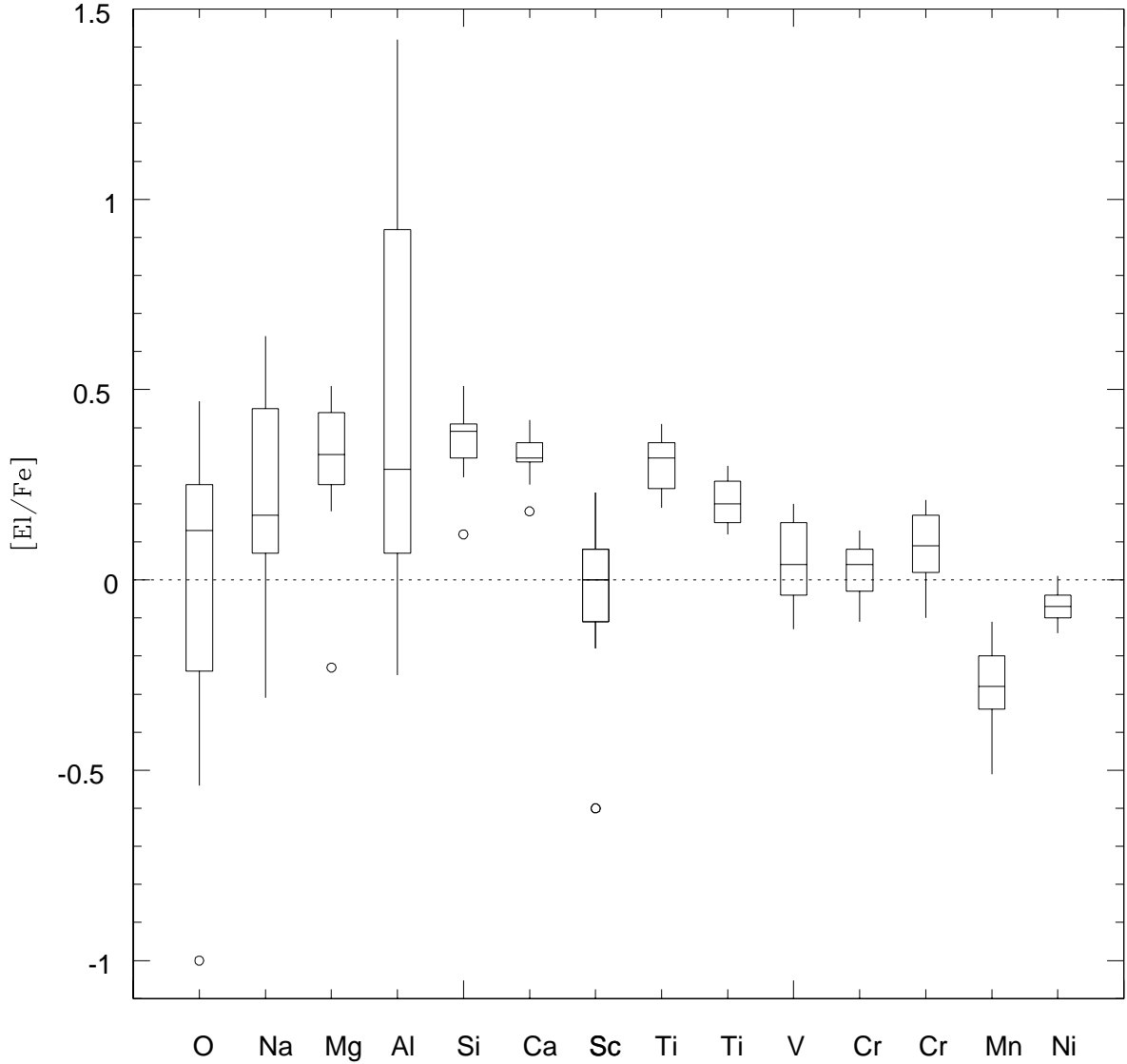}
\caption{Box-and-whiskers plot of abundance ratios in stars of the RGB in 
NGC 2808 from the present study and from \citet{car04nao28}. The line splitting
each box is the median for the corresponding element distribution. The bottom
and the top of the box are the 25th and the 75th percentiles, whose distance 
(i.e. the box length) indicate the interquartile range, encompassing the 
middle 50\% of the data. The interquartile range is used to detect outliers,
as points lying more than 1.5 times this range from the 25th and 75th
percentile. The existing outliers are indicated by open circles. The vertical
``whiskers" protruding from the box reach the maximum and minimum value of the
range excluding the outliers. The double labels for Ti and Cr indicate the
abundances from different ionization stages (first from neutral, second from
singly ionized lines).\label{f02}}
\end{figure}

The technique of using the mean and standard deviation to describe the spread of
data works at best when the distributions are symmetrical, with no outliers,
which may not be the case here.
A graphical illustration of the spread of elements in NGC 2808 is provided by
Figure~\ref{f02} which is a 
box-and-whiskers plot \citep{tuk77}, a method well suited to explore whether
a distribution is skewed and whether there are potential outliers. This kind of
plot is useful to display immediately the median and the distributions of
values in the range derived for each element in our sample, how they are
skewed, and if there are outliers.

Figure~\ref{f02} summarizes the pattern of elemental abundances 
in NGC 2808, quite similar to that of a typical globular cluster: the elements 
involved in the proton-capture processes are very spread out, the
$\alpha$-fusion elements are overabundant and the Fe-group elements are more
or less solar. 

Different groups of elements are discussed in the following. To better place
the issue of abundances in NGC 2808 in the framework of the typical pattern 
(if any does exist, see \citealt{iva01}) of clusters at [Fe/H]$\sim -1$ we will compare the
[X/Fe] ratios found in the present study with those recently obtained for 
cluster of similar metallicity from high resolution echelle spectra of
comparable quality. We then used as a comparison stars in M4 and M 5 
([Fe/H]$=-1.19$ and $-1.11$ dex on the metallicity scale
by \citealt{cg97}, a good match for the value [Fe/H]$-1.14$ derived for NGC
2808). For M 4 we used the work by \citet{iva99}, and for M 5 we adopted the
results from two different studies: \citet{ram03}, who observed stars spanning 
a large magnitude range along the RGB, and \citet{iva01}. 

We used the quoted or known solar reference abundances used in the original papers
to apply offsets to published [X/Fe] ratios, bringing them onto our own scale 
\citep{gra03}. Although systematic shifts due to the temperature scale or to 
different adopted scales of transition probabilities and (to a much lesser
extent) to the grids of model atmospheres might affect the comparison, we are
confident that the net effect cannot be very large. On the other hand, due to
the different adopted solar abundances, offsets as large as about 0.30 dex 
(e.g. between our [Al/Fe] values and those from the quoted analyses) might well
be present. These offsets were taken into account, before comparing the
different datasets.

\subsection{Proton-capture elements}

\citet{car04nao28} presented for the first time the classical Na-O
anticorrelation in NGC 2808, based on the same sample of stars analyzed here.
They show that large Na and O abundance variations are found from the RGB tip
down to the luminosity of the RGB-bump. In this respect, NGC 2808 simply joins
the growing number of clusters where this signature of $p-$capture processes
at high temperature is observed. However, it seems that stars in NGC 2808 are
able to reach very large O depletions, as also seen in M 13, the 
paradigm cluster for these chemical anomalies. For the most O-poor star (50119)
an $upper$ limit of [O/Fe]$=-1$ dex was derived from spectral synthesis of a
quite high-$S/N$ spectrum. This abundance well agrees with the minimum 
value of [O/Fe]$=-0.97$ dex (after correction to our solar reference abundances)
found for star IV-25 in M 13 in the recent study of 
\citet{sne04}\footnote{Since it is not otherwise stated, we will assume that
every O abundance in M 13 is based on actual detections, not upper limits.}.
It seems that super O-poor stars are not an unique prerogative of M 13.

However, Na enhancements appear to be somewhat larger in M 13 than in NGC 2808:
taking into account corrections for NLTE and the offset discussed above, 
a maximum of [Na/Fe]$=+0.94$ dex is measured for star M 13 IV-25,
compared with a maximum of +0.64 dex found in NGC 2808.

Mg and Al are the heaviest elements among those involved in the $p-$capture
reactions near or within the H-burning shell in cluster giants. The production
of Al at the expense of Mg $via$ the MgAl-cycle (see e.g. \citealt{lan95})
does not result into a well-defined 
anticorrelation as the Na-O one. In fact, large variations in Al are often
accompanied by much smaller changes in Mg abundances.

\begin{figure}
\epsscale{1}
\plotone{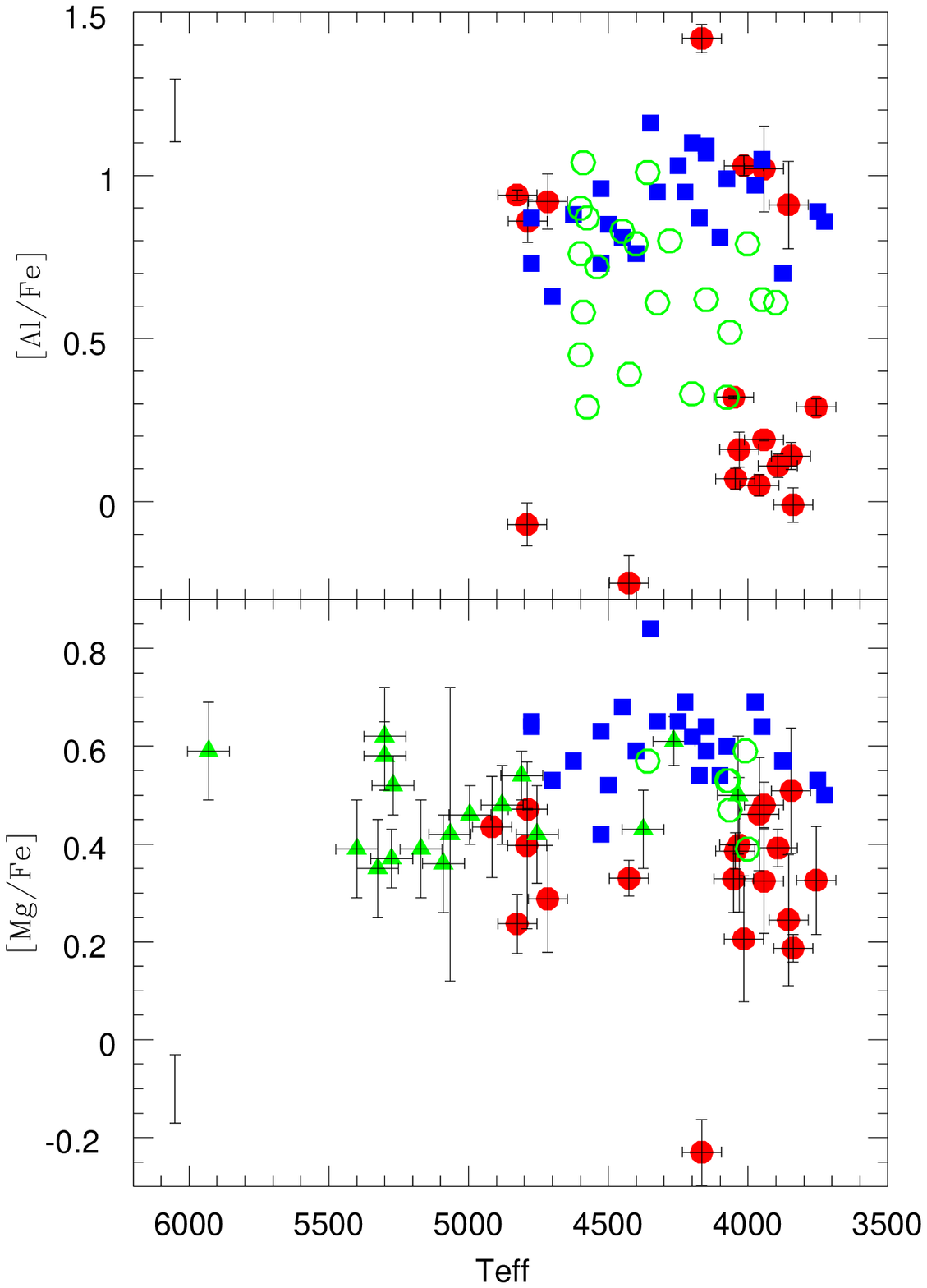}
\caption{Upper panel: run of the [Al/Fe] ratios as a function of T$_{\rm eff}$
in NGC 2808 (filled red circles;
present study), M 4 (filled blue squares; \citealt{iva99}) and M 5 (open green
circles; \citealt{iva01}).
Lower panel: run of the [Mg/Fe] abundance ratios as a function of 
T$_{\rm eff}$. Filled green triangles are stars in M 5 from \citet{ram03}. 
The error bar on
individual stars, when available, is the variance about the mean abundance in
that star, i.e. the 1$\sigma$ rms value about the mean abundance weighted for
the number of measured lines. The error bar with no data points represents the
total internal error as given in the last column in Table~\ref{tab4}, 
i.e. the expected error due
to uncertainties in the atmospheric parameters and errors in the $EW$s; this was
computed as the straight mean of total errors for the bright star 48889 and the
faintest star in our sample.\label{f03}}
\end{figure}

In Figure~\ref{f03} we display the abundance ratios of Al and Mg against 
T$_{\rm eff}$ for stars in NGC 2808, M 4 and M 5 from the present study and the
quoted papers. 
Associated to each star we plot the typical random error in T$_{\rm eff}$
and the statistical error of the mean abundance ($\sigma/\sqrt N$ where N
is the number of lines used for that star). This is a formal statistical 
uncertainty, it includes internal errors only and may be useful to give an idea
of the number of features for a given element and how reliably they could be 
measured over all the (sometime large) magnitude range of the samples.
No such errors are available for the samples in M 4 and M 5 studied by 
\citet{iva99} and \citet{iva01}.
The error bar with no associated point is the total internal error as
given by the mean of the total uncertainties found for stars 48889 and
43217 in Table~\ref{tab4}, to give an average estimate over all the 
sampled magnitude range in NGC 2808.

From Figure~\ref{f03} the distribution of Al abundances in NGC 2808 
appears to be bimodal, with two groups of stars clustered at high (about 1 dex)
and low values of [Al/Fe]. This is clearly different from the behavior found
by \citet{iva99} and \citet{iva01} in M 4 and M 5, respectively. A quite large 
spread exists in M 5, where however the distribution is rather
continuous, extending from high [A/Fe] ratios down to the level corresponding to
Al-poor stars in NGC 2808. In M 4 the Al abundances are more clustered around
[Al/Fe]$\sim 1.0$ dex. In neither clusters there is evidence of two 
distinct groupings. Unfortunately, no Al abundance was derived in M 5 by
\citet{ram03}. 

On the other hand, although we can not totally exclude that the
separation in two groups in NGC 2808 is a spurious effect due to small 
sample statistics, the gap well exceeds the $3\sigma$
internal errors. There is no correlation between Al abundances and $S/N$
values; hence,
bright and faint stars are present in both groups, suggesting that (i) the
separation is not likely to be an artifact of the analysis and (ii) there is
not strong evidence of a change or a segregation according to the evolutionary
status, at least for the 3 magnitudes below the RGB tip.
In turn, point (ii) suggests that the observed spread 
was produced in other stars, massive enough to reach the high temperature in
the H shell required to forge Al. As a consequence, we do not expect an evolutionary
pattern such as the increase of anomalies (here in the [Al/Fe] ratio) 
approaching the tip of RGB.

In the lower panel of Figure~\ref{f03} [Mg/Fe] ratios  in NGC 2808, M4 and M 5
are compared. Mg abundances do not show a clear anticorrelation with Al, and in
M 5 and NGC 2808 the average [Mg/Fe] ratio is typical of halo stars of similar
metallicity (see e.g. \citealt{gra03}); the average ratio in M 4 seems to be
slightly larger than in the other two clusters.

Star 50119 stands out with a [Mg/Fe] ratio well below the solar value; this
star is also showing the highest abundance of Al in the sample. This is hardly
surprising, since this star is the most O-poor presently observed in NGC 2808:
very low O, high Na, high Al and low Mg are the classical signature of
matter processed through the high-temperature NeNa and MgAl cycles polluting
the gas out of which the stars formed. The extreme values shown by abundance
ratios in this star suggest that a fraction of stars in NGC 2808 might have
been subjected to heavily polluted matter: among more than 60 giants in
Figure~\ref{f03}, star 50119 shows the most severe depletion in Mg, even
exceeding those observed in the template cluster M 13. 

The sum Mg$+$Al, available for 18 stars, is quite constant over all the
magnitude range. On average, log$\epsilon$(Mg+Al) = 6.68$\pm 0.02$,
$\sigma=0.07$ dex (18 stars). Notice that the observed spread is well below the
combined error of about 0.12 dex expected from the contribution of
uncertainties in atmospheric parameters and $EW$ measurement. Even the heavily
polluted star 50119 has a sum of 6.70 dex ,in very good agreement with the
average Mg+Al sum. The inescapable conclusion is that we are seeing 
a reshuffling of Al and Mg abundances in NGC 2808. 

\subsubsection{The Na-Al correlation}

Better informations can be obtained from the expected correlation of elements
such as Na and Al, predicted to be simultaneously enhanced when the NeNa and
MgAl cycles are both acting: features of these elements are in general easier
to measure and  the associated changes in abundance are usually larger than for
O and Mg.

Observing large variations in Al is a clearcut confirmation that the chemical
pattern is manufactured in stars more massive than those under scrutiny,
because presently evolving low-mass giants are unable to reach the high
temperature required to overcome the higher Coulomb barriers of heavier
elements such as Mg and Al, with respect to O and Na \citep{lan97}.

\begin{figure}
\plottwo{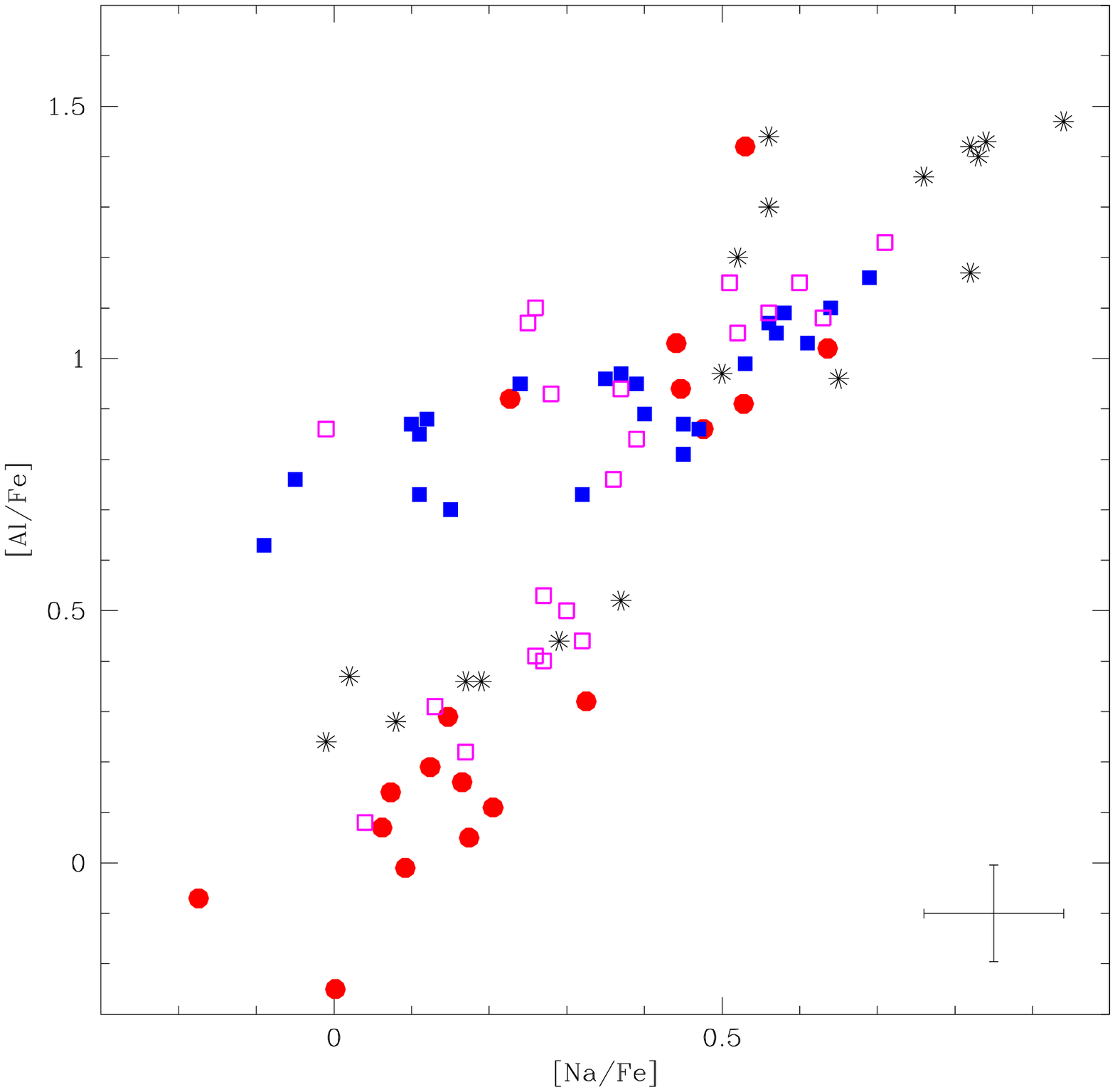}{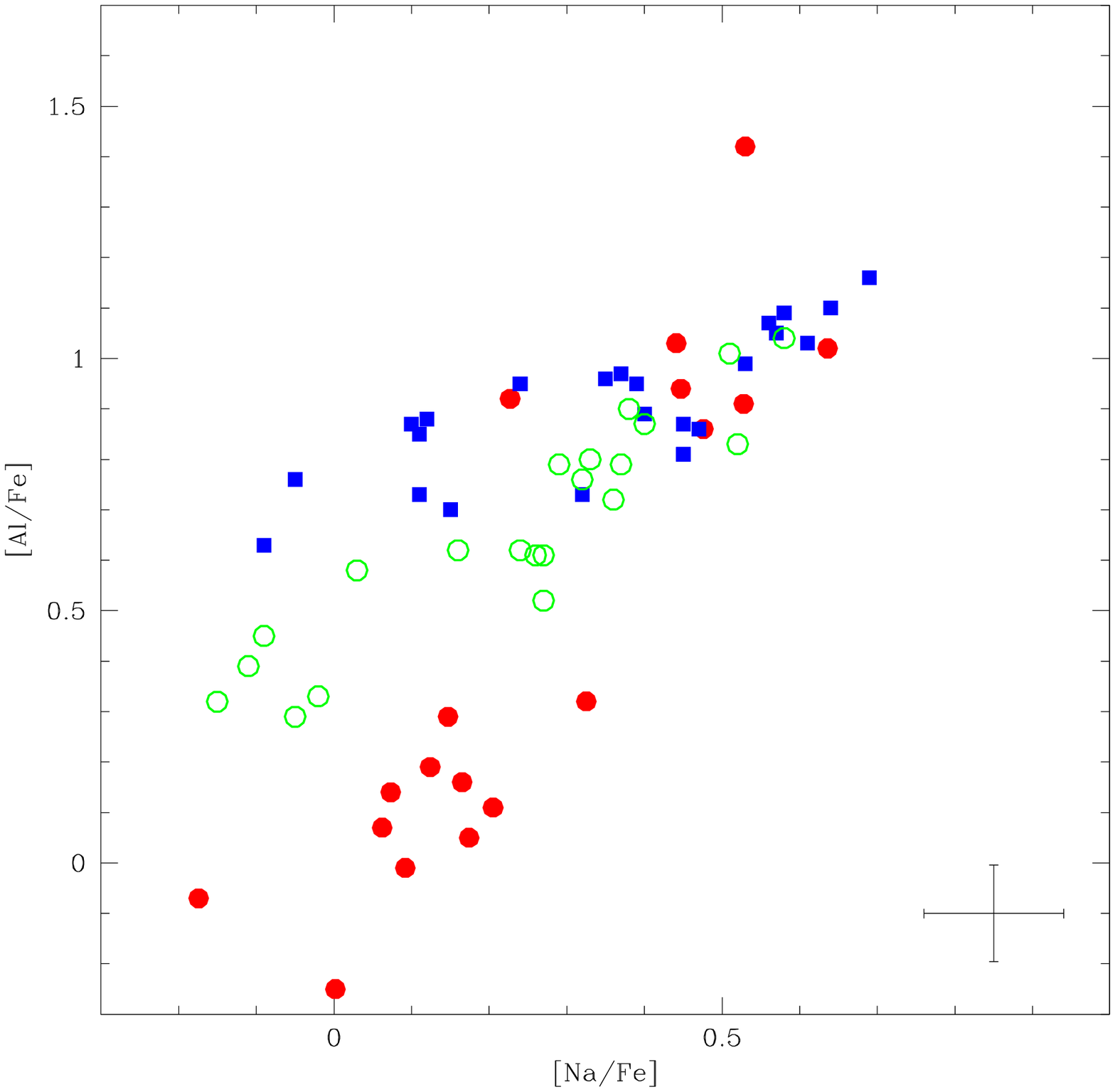}
\caption{Left panel: [Al/Fe] ratios against [Na/Fe] ratios for NGC 2808 (filled
red circles), M 4 (filled blue squares), M 3 (empty magenta squares) and M 13
(black asterisks).
Right panel: the same diagram for NGC 2808, M 5 (empty green circles) and
M 71 (black crosses)
metallicity.\label{f04}}
\end{figure}

In Figure~\ref{f04} we show the correlation between Na and Al abundances in a
number of clusters. In the left panel we plotted data for M 4, NGC 2808 and for
red giants in M 3 and M 13 from the recent work by \citet{sne04}. Although
slightly more metal-poor than the other clusters ([Fe/H]$\sim -1.35$ dex on the
\citealt{cg97} scale), these two other datasets help to highlight a possible 
feature in these diagram, maybe overlooked before.  We tentatively put forward
a working hypothesis: that what we are seeing is not a scattered plot, but a
two-branches diagram. In other words, while a single set of data would have the
appearance of a single Na-Al correlation with some scatter, the superposition
of several clusters would seem to define two main loci, with differently 
sloped correlations. In the lower branch, from [Na/Fe],[Al/Fe]$\sim$0.0,0.0 to
0.4,0.8, a change of about 0.5 dex in Na corresponds to a variation of about
0.8 dex in Al; in the upper branch (from $\sim -0.1,0.7$ to about 0.5,1.0), the
slope is much shallower and 0.5 dex in [Na/Fe] seems to translate into a change
of only  0.3 dex in [Al/Fe]. At about 0.5,1.0 the two branches intercept with
each other.

The various clusters seem to differently populate these two putative lower and
upper branches in what we can call a ``LU-plot" (L and U refer to lower and
upper branch, respectively). Stars of NGC 2808 are
spread mainly  along the L-branch, with some stars defining a stubby U-branch;
the same occurs for M 3, even if its U-branch seems to be more
extended and defined than in NGC 2808. However, stars in M 13 (the 
second-parameter twin 
of M 3) populate $only$ the L-branch, no one is falling along or near the
U-branch. On the other hand, $all$ stars in M 4 seem to exclusively populate the
U-branch or its prosecution to higher Na,Al values.

To strengthen the reliability of this LU-plot, we note that abundances for both
M 3 and M 13 come from the very same study and a highly homogeneous 
procedure;
hence, it would seem rather difficult to think of some effects in the 
abundance derivation able to produce a scatter in [Na/Fe] vs [Al/Fe] $only$ for
some giants in M 3. In fact about 0.5 dex in Na abundances would be required to
force all the M 3 sample to lie along a single and well defined relation as for 
M~13.

For sake of clarity, in the right panel of Figure~\ref{f04} we plotted
again for reference the data for NGC 2808 and M 4, and we superimposed the
Na-Al relation for M 5.
The case of M 5 seems more controversial, since its stars appear to fall in
between the two branches and with a slope intermediate between those of M 4 and
NGC 2808.

However, additional offsets
might be needed, depending on whether the corrections for NLTE in Na are taken 
into account or not in the original studies.
In their last analysis of M 3 and M 13 \citet{sne04}
included the corrections, based on our same prescriptions \citep{gra99}; however,
neither M 5 nor M 4 were treated in this way, and an additional offset in
[Na/Fe] (estimated at about +0.1 dex in this metallicity and temperature range)
could be required for these two clusters. This would shift stars in M 5 
nearer to the L-branch, at least partly eliminating the M 5 discrepant
behavior.

\begin{figure}
\epsscale{1.1}
\plotone{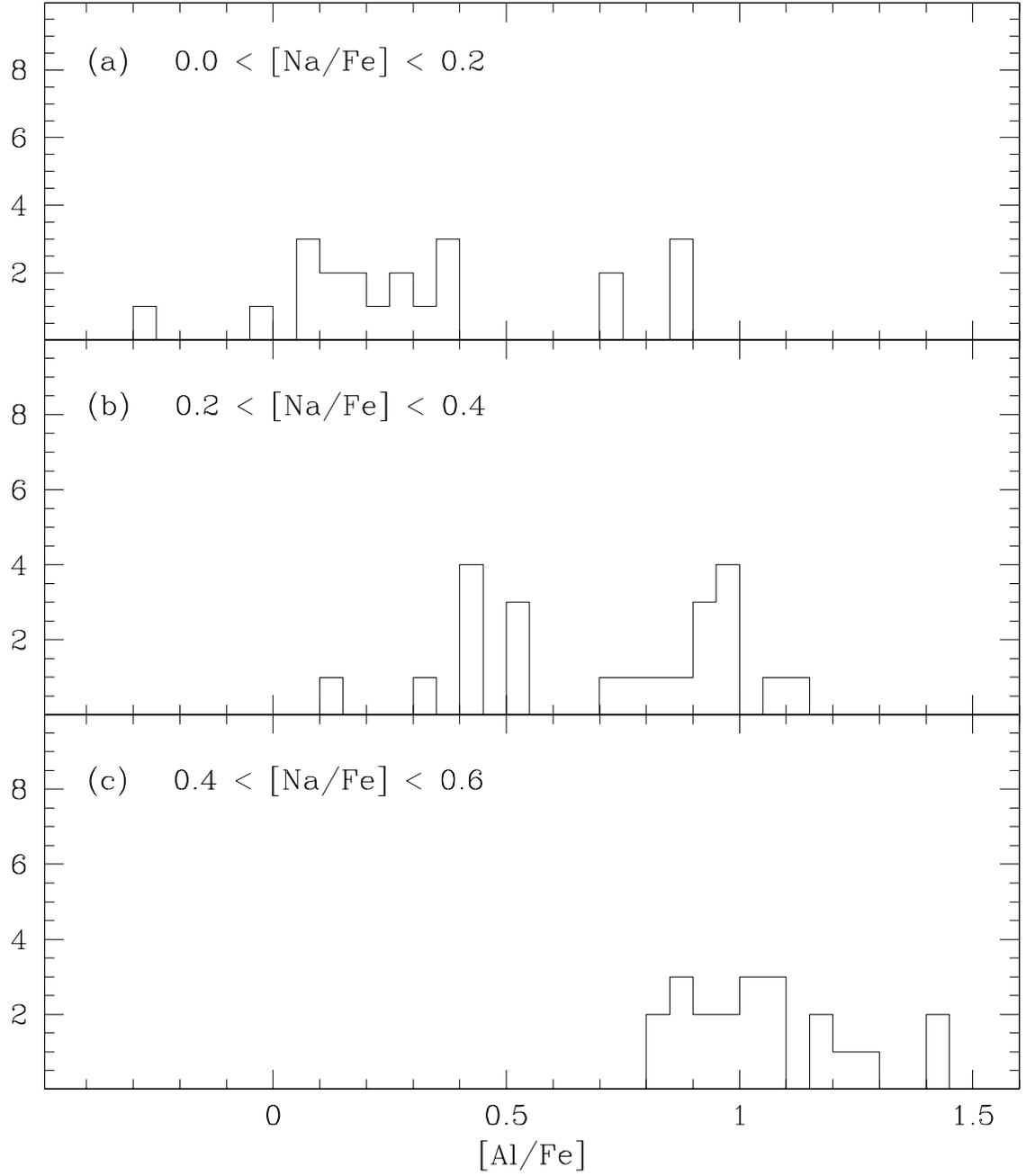}
\caption{Distributions of [Al/Fe] ratios for stars plotted in the left panel of
Figure~\ref{f04} in three ranges of [Na/Fe] values, shown in each 
panel.\label{f05}}
\end{figure}

To have a deeper insight into the LU-plot we divided the stars plotted in the
left panel of Figure~\ref{f04} into sub-samples, considering three slices
in [Na/Fe] ratios: (a) $0.0 < $[Na/Fe]$ < 0.2$ dex, 
(b) $0.2 < $[Na/Fe]$ < 0.4$ dex and (c) $0.4 < $[Na/Fe]$ < 0.6$ dex. The
corresponding distributions in [Al/Fe] are plotted in Figure~\ref{f05}. This
Figure clearly shows how the gap in [Al/Fe] between the L and U branches of the
LU-plot decreases as the two branches approach the intersection point.
To investigate the statistical significance of the bimodality we performed a
Kolmogorov-Smirnov test: the probability that distributions (a), where the 
gap in the LU-plot is most prominent, and (b), where the two branches merge,
are extracted from the same parent population is $8 \times 10^{-8}$. This
probability decreases to 0.003 when comparing distributions (b) and (c).
Finally, the probability for distributions (a) and (b) being extracted from the
same parent population is only $5 \times 10^{-5}$.

A full re-analysis of literature results is beyond the purpose of the present
study; anyway, regardless of the precise attribution of clusters to either one
of the LU-branches, what seems to be a rather robust conclusion is that the 
slope in
the Na-Al relation is not unique. In M 4 {\it it is} shallower than in 
M 13 and NGC 2808, and in M 5 it $could$ be somewhat steeper than in M 4.
We also note that different slopes and intercepts were noticed between the Na-O
anticorrelations in NGC 288 and NGC 362 by \citet{shet00}.

If this result will be confirmed by more extended databases and more
homogeneous analyses, what are the implications? Is there a change in the
rates or efficiencies of the NeNa and MgAl cycles related to some cluster properties? The
global metallicity of M 5, M 4 and NGC 2808 is virtually the same, yet the
Na-Al correlation is differently sloped. 
The forging of Na and Al in $p-$capture reactions requires different
temperatures \citep{lan95,lan97}: do the different slopes observed imply a different temperature
profile in the interior of stars that contributed the yields? In turn, this
could imply a different range of the involved masses.
Although this is only a working hypothesis, a real difference in the slope of
the Na-Al correlation, if confirmed, could provide precious constraints to the
theoretical modeling of the input physics required to fully explain the 
nucleosynthesis products seen in cluster stars.

\subsection{$\alpha-$elements}

\begin{figure}
\epsscale{1.1}
\plotone{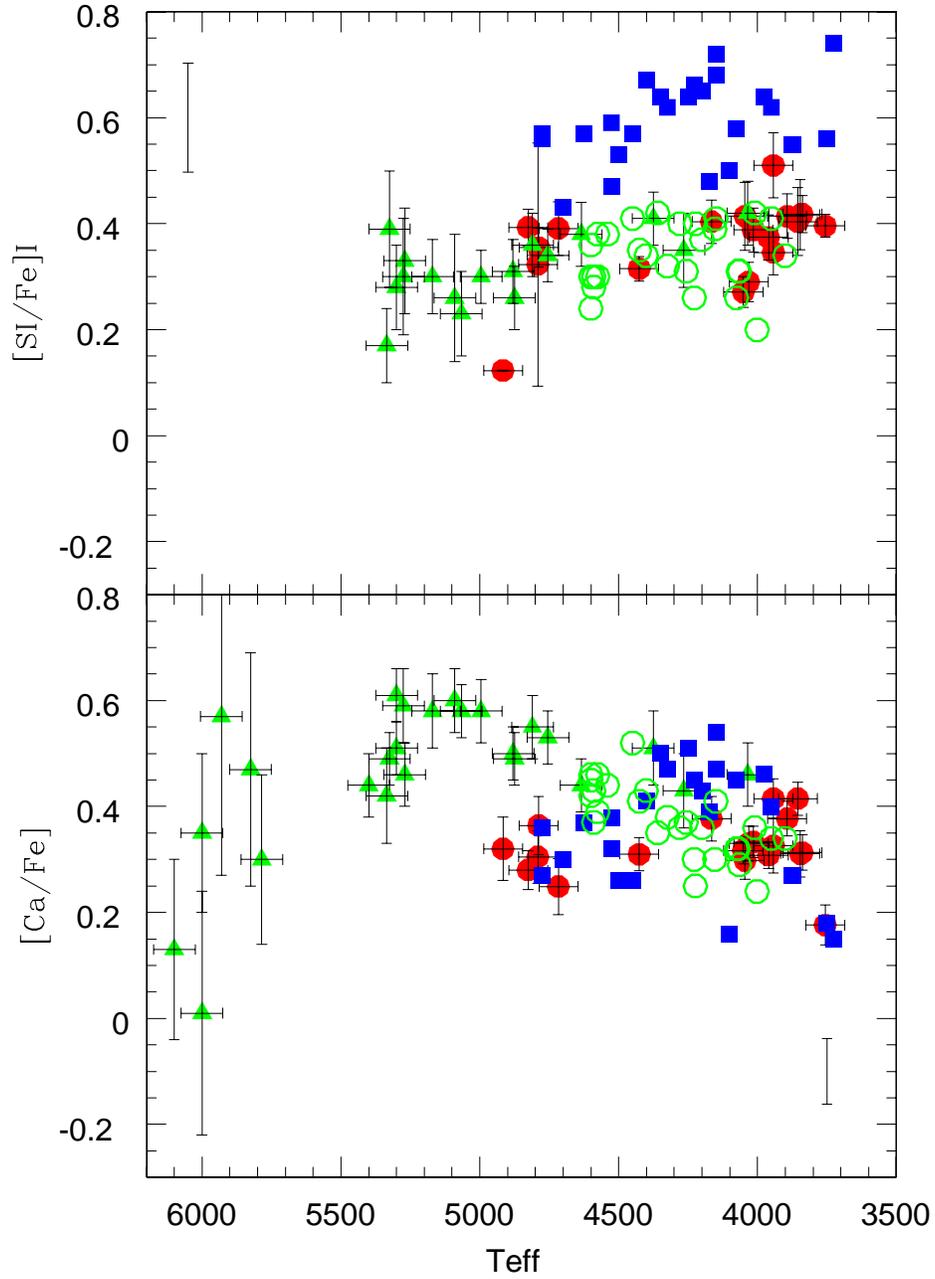}
\caption{Run of the [Si/Fe] (upper panel) and [Ca/Fe] ratios (lower panel) 
as a function of the effective temperatures in globular cluster red giants.  
Symbols are as in the previous Figure~\ref{f03}.\label{f06}}
\end{figure}

\begin{figure}
\epsscale{1.1}
\plotone{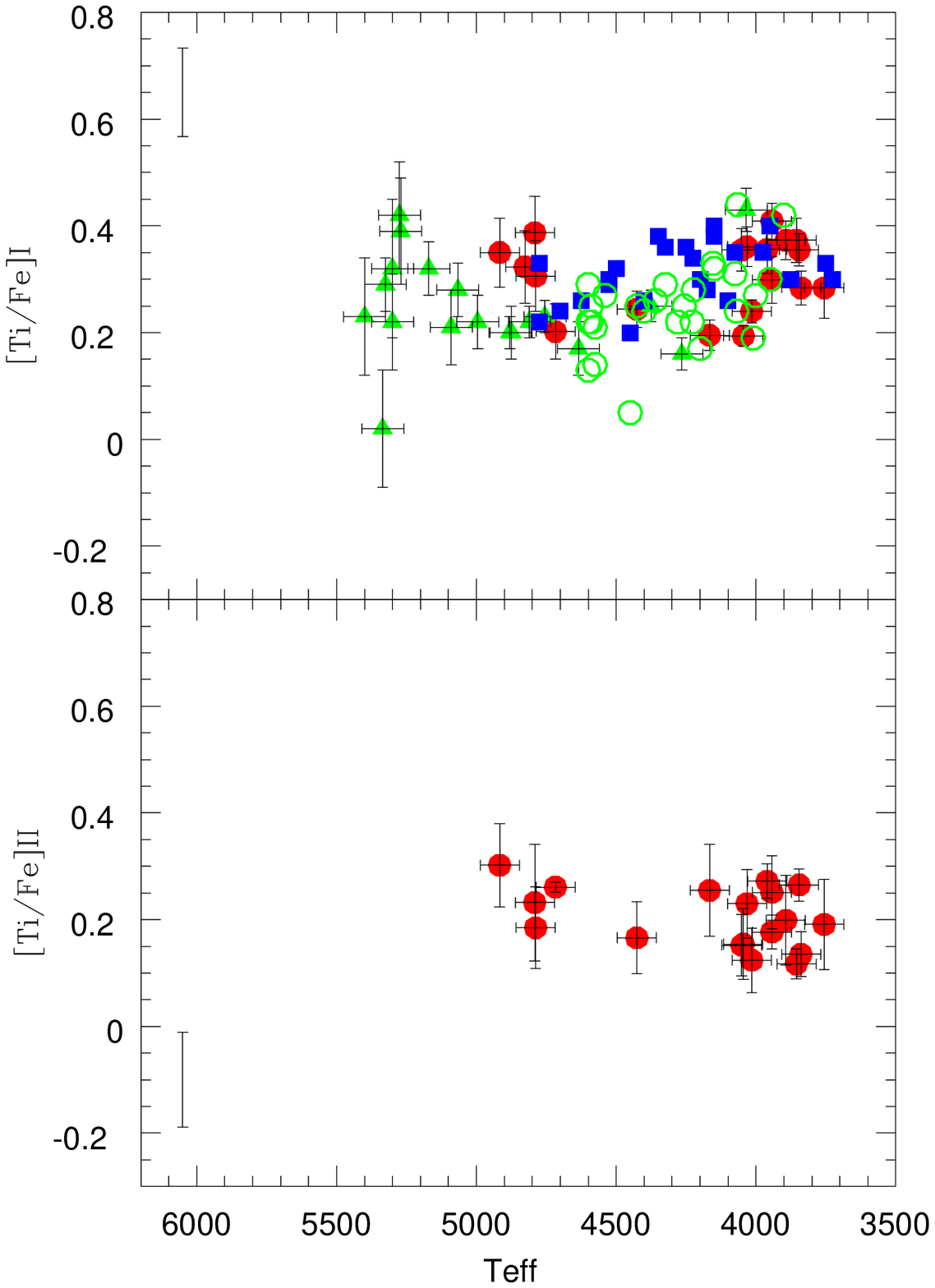}
\caption{Run of the [Ti/Fe]I (upper panel) and [Ti/Fe]II ratios as a function
of T$_{\rm eff}$. Symbols are as in previous Figure.\label{f07}}
\end{figure}

Abundances of $\alpha-$process elements in stars of NGC 2808 are shown in 
Figure~\ref{f06} and Figure~\ref{f07}. The element ratios show 
the normal overabundance above the solar value typical of
the halo stars of similar metallicity. When compared to the average
overabundances found in field stars of the dissipative component of our Galaxy
\citep{gra03} with [Fe/H] in a $\pm 0.2$ dex range centered on the metallicity
of NGC 2808, the mean values in Table~\ref{tab5} for Si, Ca, Ti I and Ti II
differ by only a few hundredths of dex. The only exception is Mg, for which we
found an average difference (NGC 2808 - field) = -0.13 dex: this result agrees
with the Mg being somewhat reduced in NGC 2808, due to the depletion resulting
in large Al enhancements.

After correcting for the different solar reference abundances, we confirm the
finding by \citet{iva99} of an unusual overabundance of [Si/Fe] in M 4, at
least with respect to other clusters of very similar metallicity.
The other $\alpha-$elements seems to be at the same level in M 5, M 4 and 
NGC~2808; however, Ca abundances in NGC 2808 present a more reduced scatter
than in the other two clusters.
The average of $\alpha-$elements (Mg+Si+Ca+Ti I+Ti II) we found in NGC 2808 
is $<[\alpha$/Fe]$> = +0.32 \pm 0.01$, ($\sigma=0.04$ dex, 19 stars).

\subsection{Fe-group elements}

\begin{figure}
\epsscale{1.1}
\plotone{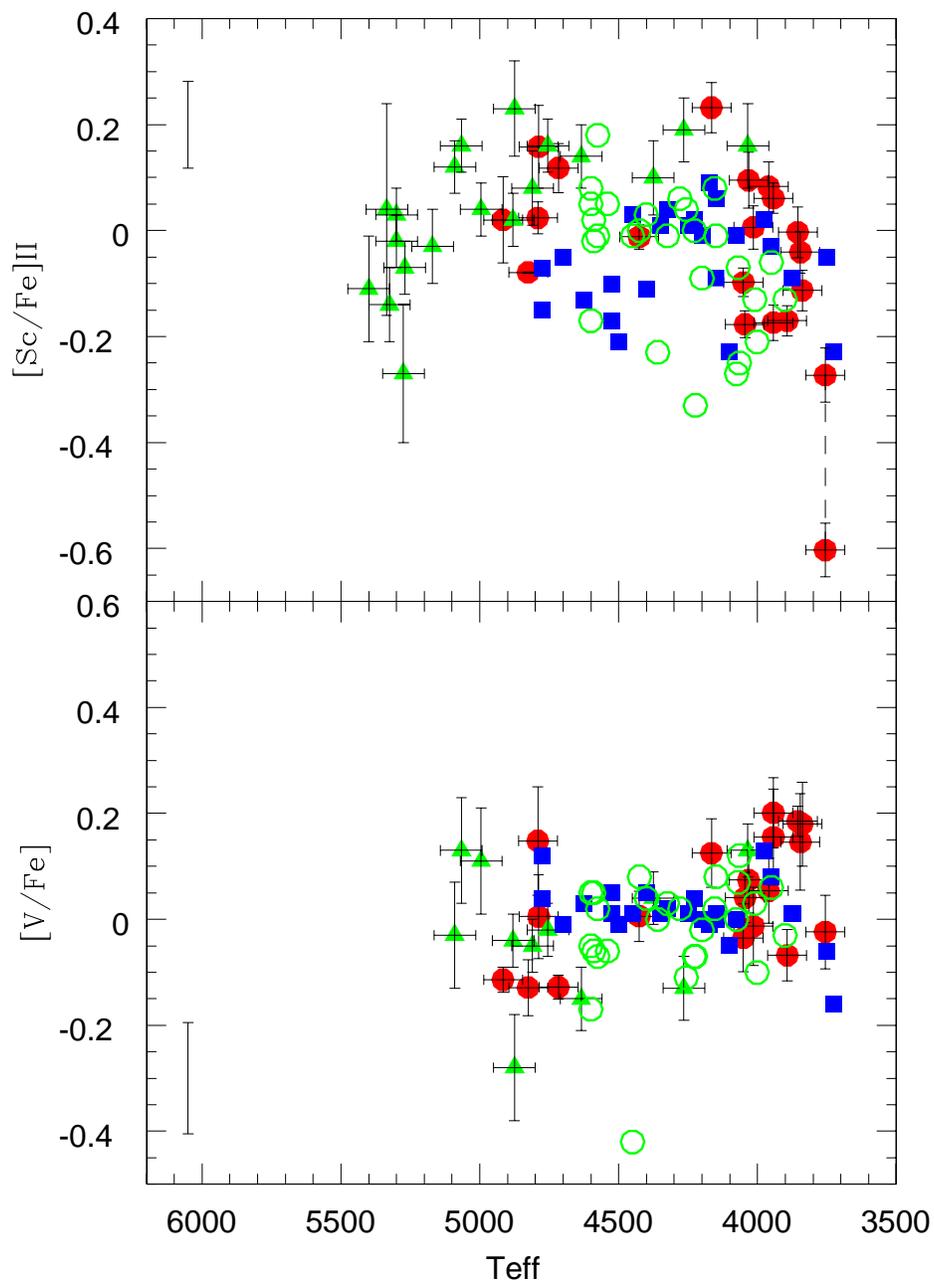}
\caption{Run of the [Sc/Fe]II (upper panel) and [V/Fe]I ratios as a function
of T$_{\rm eff}$. Symbols are as in previous Figures. The dashed line connects
the ratio [Sc/Fe]II for star 50761 of NGC 2808 (the coolest in the sample) as 
computed 
with the actual [Fe/H]II value for this star (lower point) and with 
[Fe/H]II$=-1.14$, average value for the whole sample (upper 
point).\label{f08}}
\end{figure}

\begin{figure}
\epsscale{1.1}
\plotone{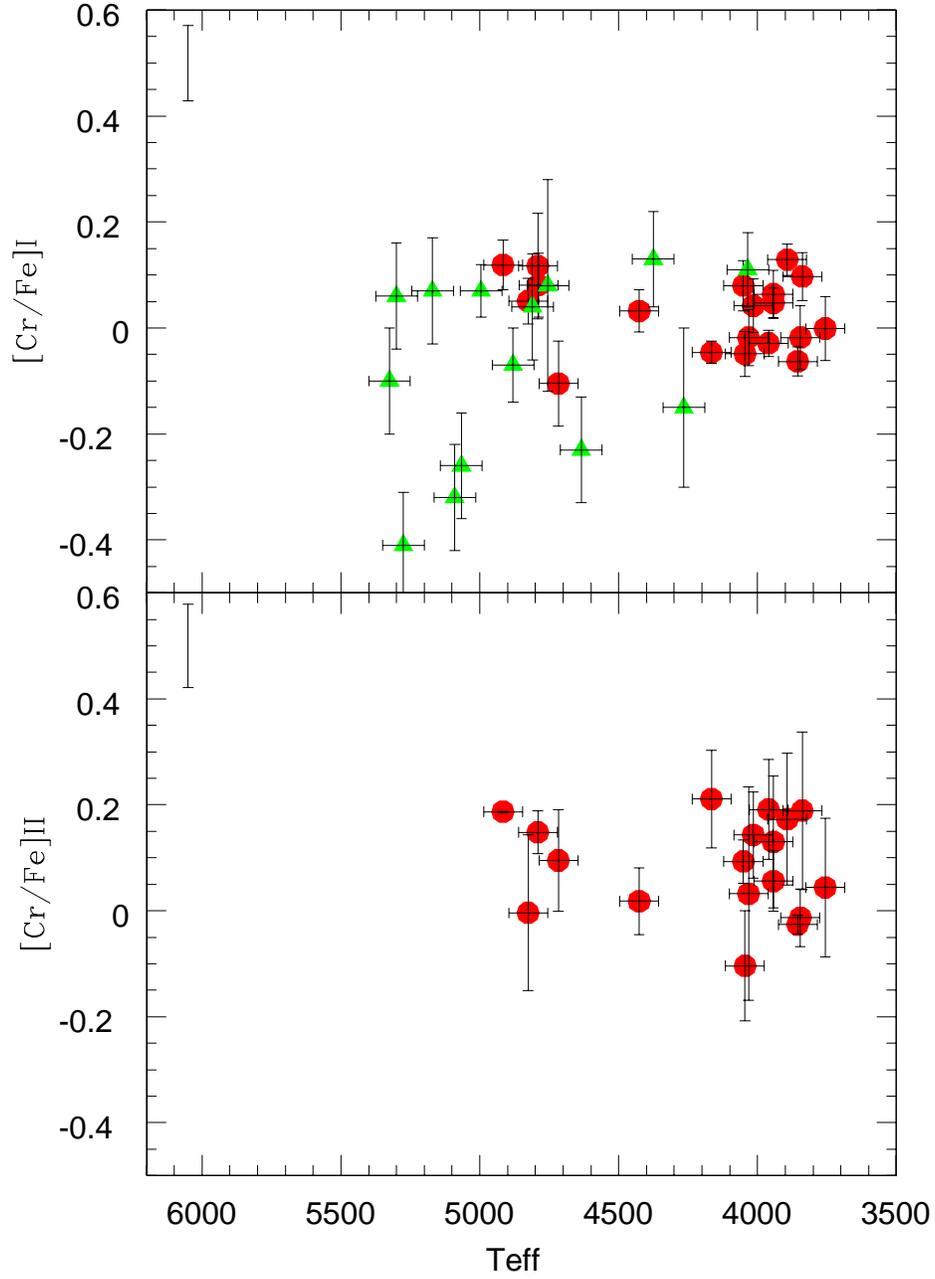}
\caption{Run of the [Cr/Fe]I (upper panel) and [Cr/Fe]II ratios as a function
of T$_{\rm eff}$. Symbols are as in previous Figures.\label{f09}}
\end{figure}

\begin{figure}
\epsscale{1.1}
\plotone{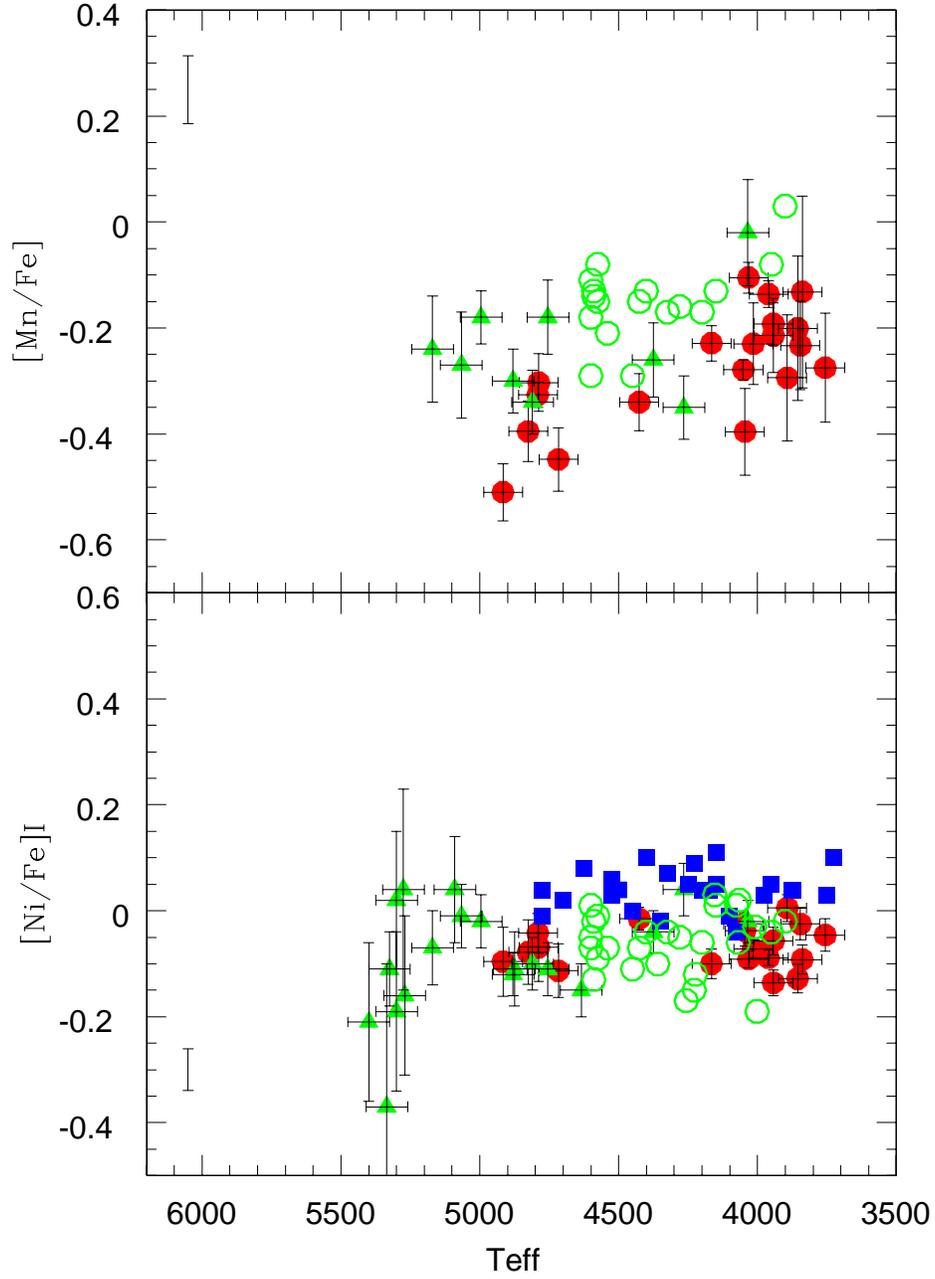}
\caption{Run of the [Mn/Fe] (upper panel) and [Ni/Fe] ratios as a function
of T$_{\rm eff}$. Symbols are as in previous Figures.\label{f10}}
\end{figure}

Figure~\ref{f08}, Figure~\ref{f09} and Figure~\ref{f10} show
the run of elements in the Fe-group for NGC 2808, as compared to the values 
in M 4 and M 5.

For Sc II the dashed line in Figure~\ref{f08} connects two values of the
[Sc/Fe]II ratio for star 50761 in NGC 2808: the lower value was computed 
using the [Fe/H]II value derived for this star, and
the upper one using the average value, [Fe/H]II$=-1.14$, 
found for NGC 2808. Since this star is the coolest in the sample, its
abundance of FeII could be not completely reliable.

We do not consider the scatter in this Figure as very significant, since in
addition to the possible offsets due to differences in the model atmosphere 
grids, temperature and $gf$ scales and 
solar analysis there is the further uncertainty due to the corrections for
the hyperfine structure, adopted from different sources in different studies.

Apart from Mn, the Fe-group elements have typical average ratios near 
the solar value.

\section{Hunting for correlations between pollution and cluster parameters}

The present study allows to add another point to the growing set of
clusters with modern abundance analysis based on high resolution spectra of a
statistically significant number of stars (N$_{\rm stars} \gsim 20$) per
cluster. The lack of consistent sets of abundances for elements involved in
$p-$capture reactions has been one major limit in searching for correlations
between the degree of 
chemical anomalies and other global globular cluster parameters. In turn, 
these links, if
found, can tell us a lot about the first billion years in the history of the
cluster itself.

To explore this long standing issue we selected a number of clusters with
recent abundance analysis for the key elements O, Na, Mg and Al:  to the 
already mentioned studies for M 4, M 5, M 3 and M 13, we added NGC 6752
\citep{yong03}, NGC 6838 (M 71, \citealt{ram02}) and NGC 7078 (M 15,
\citealt{sne97}). This allows to well sample the whole metallicity
range ($-2.5<$[Fe/H]$<-0.5$) covered by the bulk of galactic globular clusters.
Note that this represents our only selection criterion, apart from 
availability of adequate quality abundance analysis of red giant stars.

An additional problem was to find a good estimator of the amount of chemical
anomalies in each cluster. We believe that lack of an objective criterion
has somewhat hampered fruitful researches in this direction up to now.
\citet{car96} explored possible connections with the HB morphology using
$\Delta$[O/Fe], the difference between the maximum and minimum [O/Fe] ratio
observed in a given cluster; however, these values might be affected by 
small numbers statistics, even
in clusters with a fair number of sampled stars.

Hence, we propose here that the interquartile range (IQR) of a distribution of
abundance ratios is not only a good graphical representation of the spread, but 
also an optimal tool to $quantitatively$ define the extension of chemical
inhomogeneities within a cluster and to compare it with other clusters.
In fact, as mentioned above,  box-and-whisker plots are ideal for comparing
distributions because the center, spread and overall range are immediately
apparent. The IQR can be used as a measure of how much spread-out the values
are and it  is less subject to sampling fluctuations in highly skewed
distributions. 

As before, the abundance ratios [O/Fe], [Na/Fe], [Mg/Fe] and [Al/fe] were
shifted onto our set of solar reference abundances and we derived values of
median and IQR for each element. Inclusion of corrections for NLTE
in Na abundances only affects the relative values of medians, not the IQRs.

In Table~\ref{tab6} we list the derived values for O, Na, Mg and Al, and
the ratios [O/Na] and [Mg/Al]: in other words, these are the most
objective measure possible of the spread along the Na-O and
the Mg-Al anticorrelations in each cluster.
The median value and IQR for [Mg/Al] in M 5 are more uncertain,
because they are based on only a few stars: \citet{ram03} did not have any Al
features in their study and only a few [Mg/Fe] ratios were
derived by \citet{iva01}.

Armed with this set of diagnostics, we can now check for possible
correlations with chemistry, structural parameters and, for the first time to
our knowledge, orbital parameters of the clusters.

\subsection{Chemistry}

\begin{figure}
\plottwo{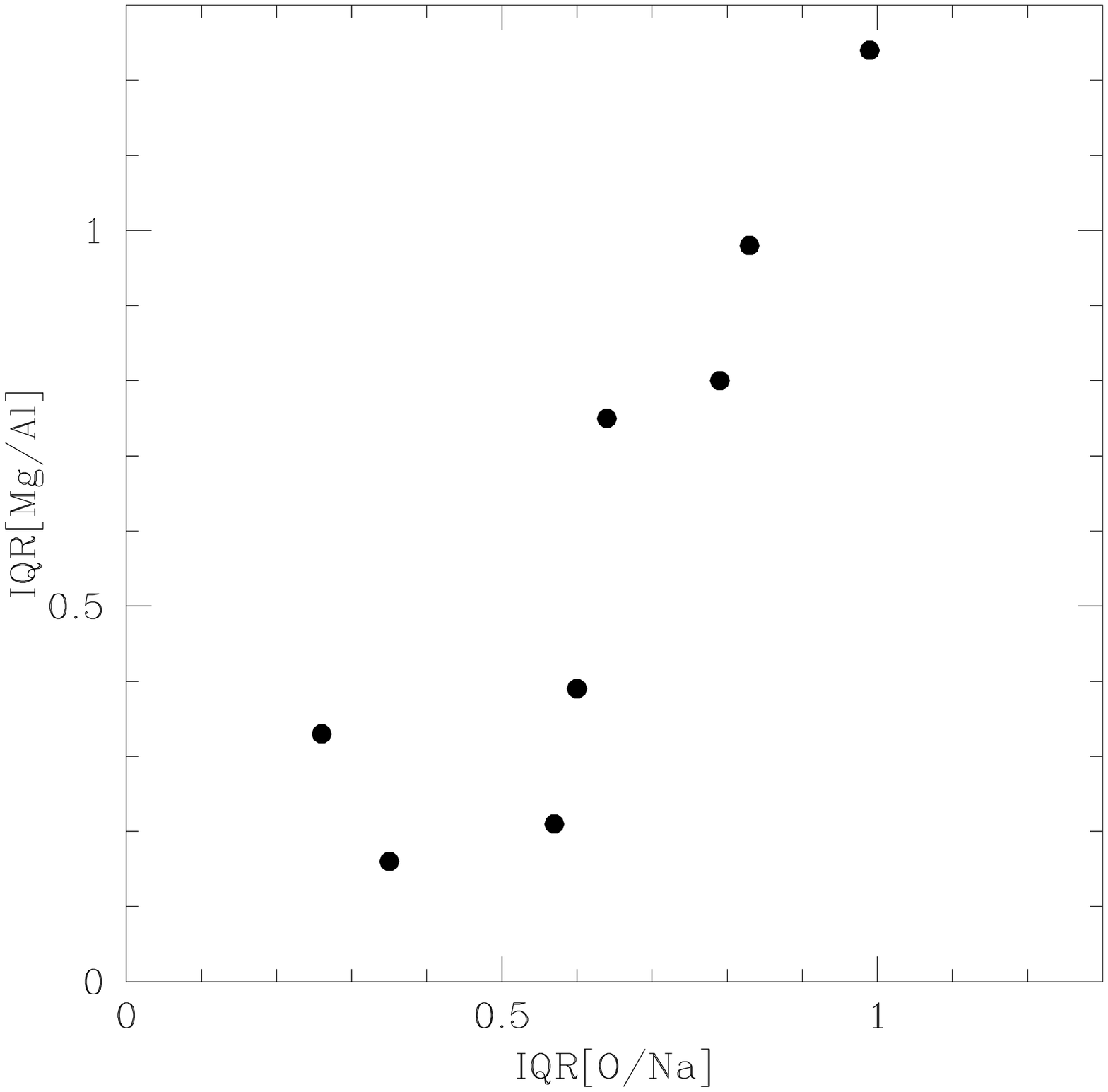}{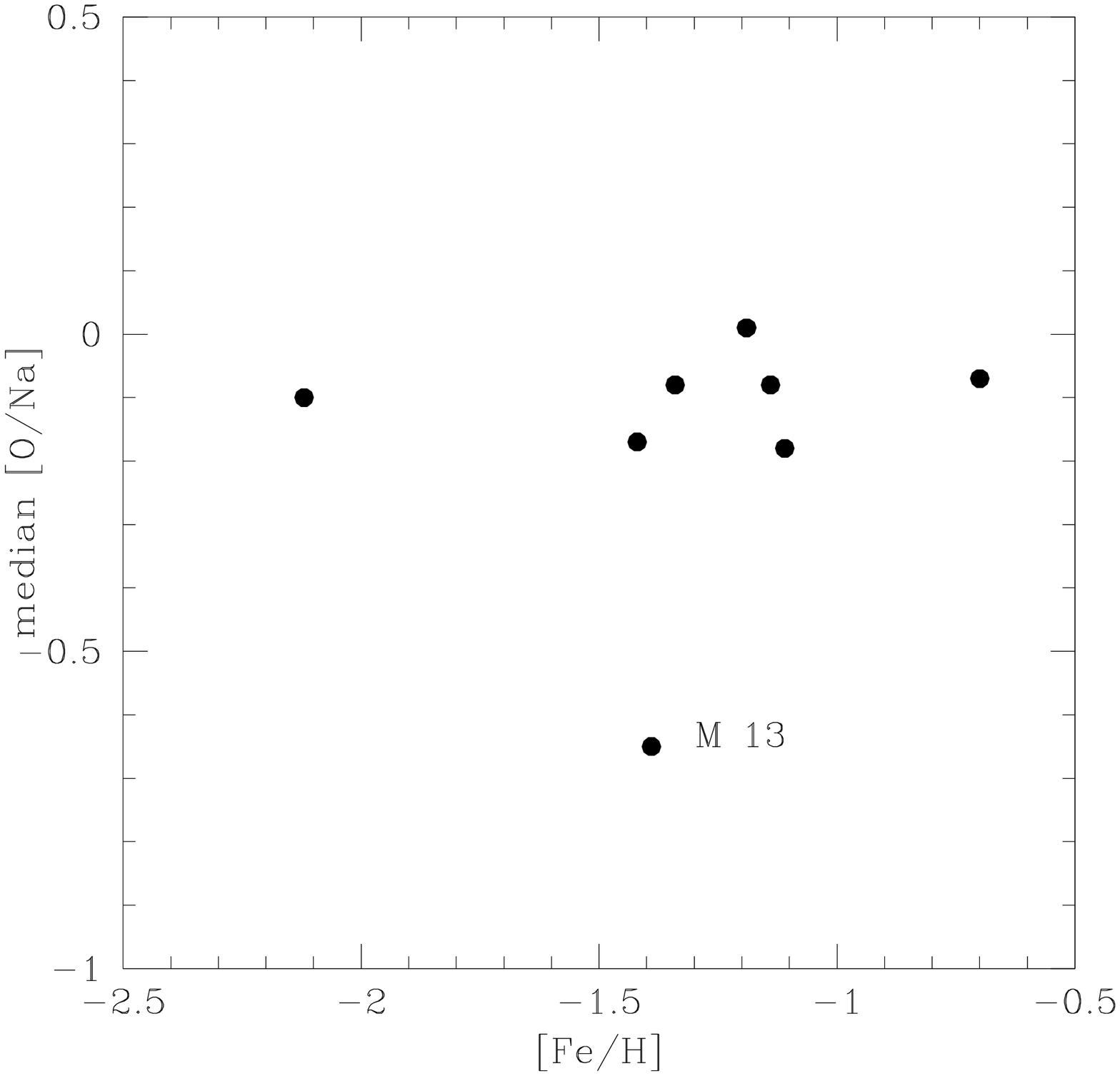}
\caption{Left panel: interquartile ranges (IQRs) along the Mg-Al
anticorrelation as a function of IQRs along the Na-O anticorrelation. Right
panel: median values along the Na-O anticorrelation against 
metallicity.\label{f11}}
\end{figure}

A first indication about how much useful the IQRs are is provided by the left
panel in Figure~\ref{f11}, where we plot the IQR along the Mg-Al
anticorrelation against those along the Na-O anticorrelation. Despite 
the fact that in each
cluster the Na-O relation is much more well defined than the Mg-Al one, these
two quantities are quite well correlated: the spread in
[Mg/Al] is increasing as the spread in [O/Na] increases. 

Although it has long been known \citep{den89,lan93} that these
element are linked in the $p-$capture chains of reactions, this quite good
correlation is a clearcut evidence that NeNa and MgAl
cycles involved are likely coming from the very same source.
In turn, since the MgAl cycle can work only at temperatures of T$_9 = 0.040$ or
larger \citep{lan93}, unreachable in low mass RGB stars, this figure 
confirms without
ambiguity that the bulk of inhomogeneities are established in conditions only
possible in the interiors of more massive stars (see \citealt{car05} for a more
thorough discussion on different classes of candidate polluters and associated
problems).
Evolutionary effects, if any, might be present only a secondary noise
superimposed to the output from these massive and more short-living stars.

In particular, the only cluster where evolutionary alterations seem to be on
average quite important is M 13 \citep{pil96}. Indeed, in other clusters an 
opposite trend can be seen (for instance in M 5, \citealt{iva01}).
In the right panel of
Figure~\ref{f11} we display the median values along the Na-O
anticorrelation as a function of metallicity. All objects are
clustered around an average value $<$median[O/Na]$> = -0.10$, $\sigma=0.06$
dex: the only one that stands well out is M 13, off by more than
8$\sigma$ from the others. It appears that M 13 is the real exception among all
clusters considered here.
Although also in NGC 2808 we found stars with O abundances as low as in the super
O-poor stars in M 13, it seems that in the latter the pollution
was able to produce a median value of [O/Na] well below those of typical
globular clusters: the main difference lies not in the minimum [O/Fe] or
maximum [Na/Fe] ratios reachable, but in the spread along the Na-O
anticorrelation. 
While this is hardly a new result, it is now on more quantitative grounds,
showing the utility of using the IQR as a new indicator.

At present, however, the question on why M 13 is so unique among its fellows
has still to remain unanswered.

Going back to Table~\ref{tab6} we can see that the IQRs of individual [X/Fe]
are also correlated with each other, although the relations involving
IQR[Mg/Fe] are somewhat shallower than the others.

Finally, no correlation whatsoever is found between IQR[O/Na],IQR[Mg/Al] and
either metallicity or age indicators: it appears that whatever the epoch of
formation is, globular clusters formed with the pattern of anticorrelated 
Na-O and Mg-Al already established early on. This in turn suggests that we are
probably dealing with something intimately related to the very same mechanism
of cluster formation.

\subsection{Pollution and structural parameters}

In the previous Section we confirmed and gave new quantitative strength to the
concept that the bulk of chemical inhomogeneities is from a (likely very early)
pollution. We need to test now whether the amount of this pollution can be related
to the structural parameters of clusters.

\begin{figure}
\epsscale{1.0}
\plotone{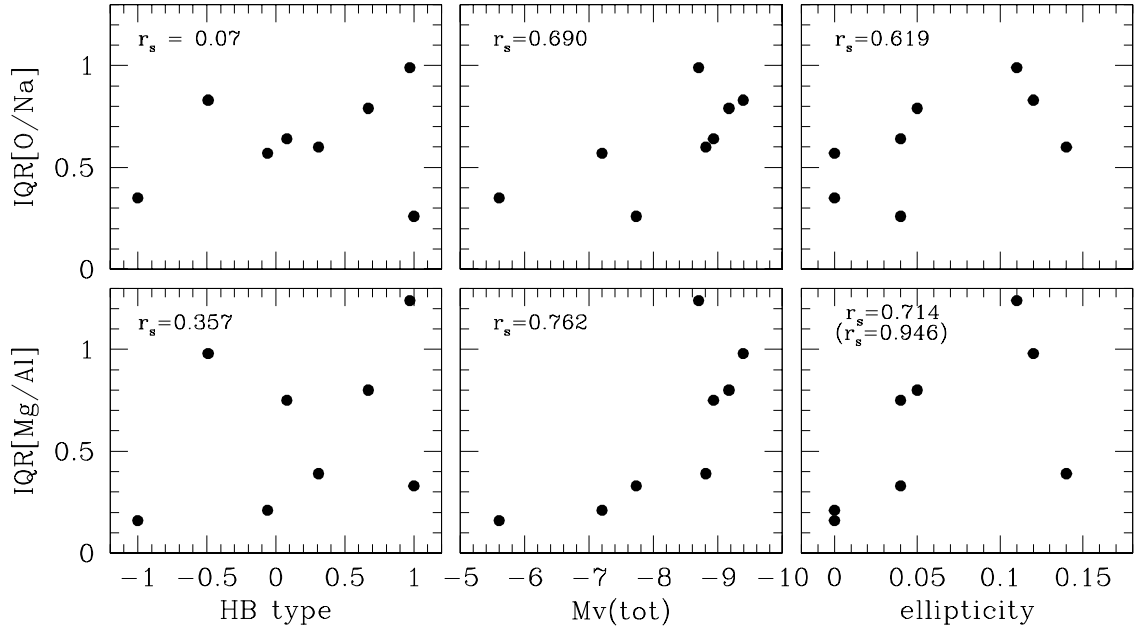}
\caption{IQRs along the Na-O and the Mg-Al anticorrelations as a function of
the HB type, total absolute magnitude and ellipticity of the 
cluster. In each panel is reported the Spearman rank correlation coefficient; 
values in brackets refer to the correlations computed excluding the 
cluster M 5.\label{f12}}
\end{figure}

\begin{figure}
\epsscale{1.0}
\plotone{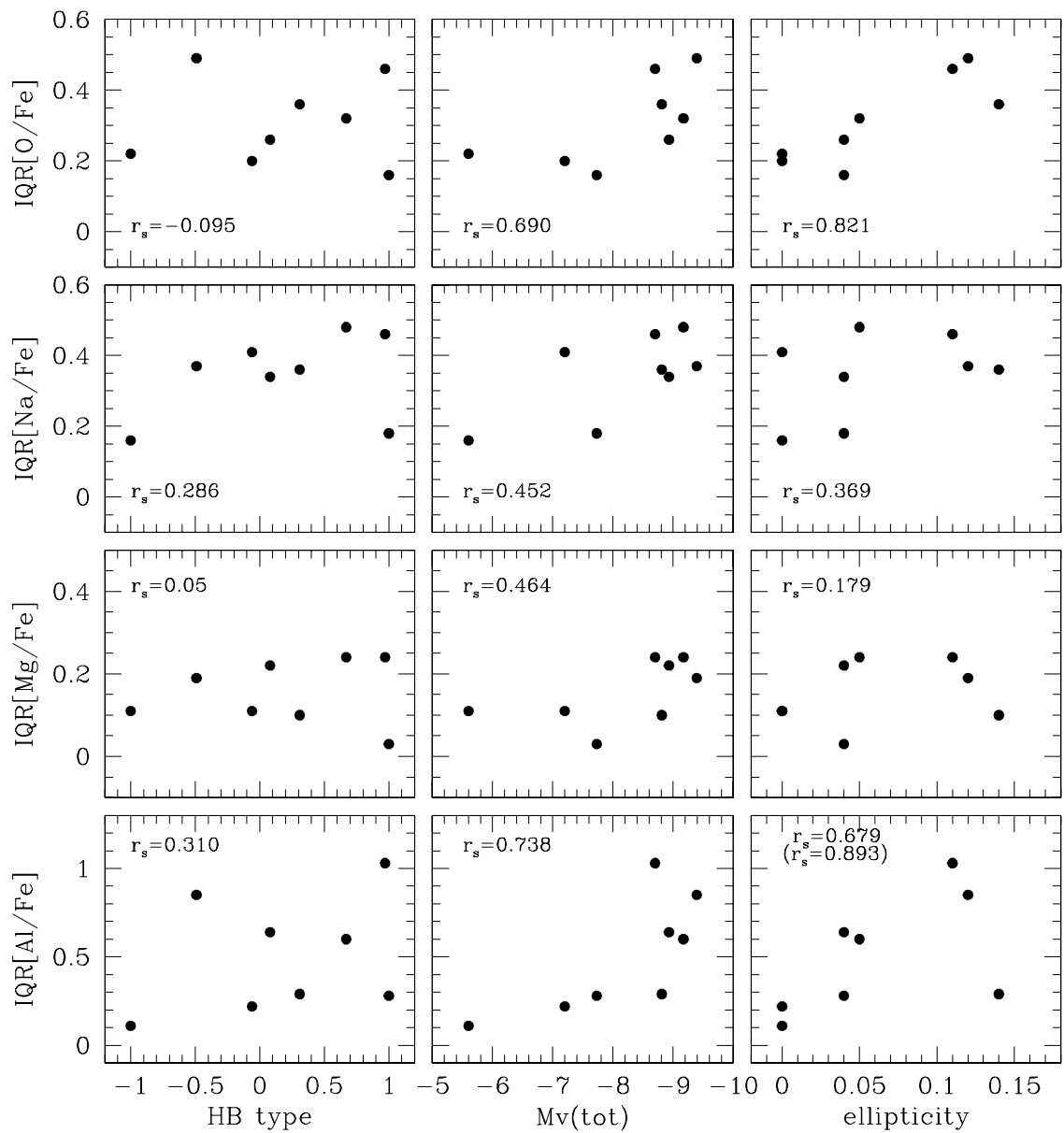}
\caption{IQRs of [O/Fe], [Na/Fe], [Mg/Fe] and [Al/Fe] distributions 
as a function of
the HB type, total absolute magnitude and ellipticity of the 
cluster. In each panel is shown the Spearman rank correlation 
coefficient.\label{f13}}
\end{figure}

In Figure~\ref{f12} and Figure~\ref{f13} we plot the IQRs of distribution along
the Na-O, Mg-Al anticorrelations and those of individual [X/Fe] ratios as a
function of the morphological HB type, the absolute visual magnitude and the
ellipticity of the clusters, taken from the updated catalog  by \citet{har96}.
To test the statistical significance (or its lack of significance) of these
correlations, we computed the Spearman rank correlation coefficient, shown in
each panel; if a value is in brackets, it refers to the  same correlation, but
obtained after exclusion of M 5 (see below).

The spread in the various distributions does not seem to depend on the HB type. 
\citet{iva01} noted a trend for clusters having a red HB morphology to be
segregated from those having a more marked blue HB morphology, as far as the
pattern along the Na-O anticorrelation was concerned. However, from the plots
in Figure~\ref{f12} and Figure~\ref{f13} we cannot confirm any strong
dependence of chemical inhomogeneities on HB type.
If the chemical composition on the RGB has some influence on the distribution
function of stars onto the HB, it has to be modulated by some other factors,
since clusters with very different HB morphologies present the same amount of
inhomogeneities. 

More promising is the run of IQRs as a function of the present-day 
cluster mass, as 
represented by the total absolute magnitude: there is a trend for the spread to
increase with the cluster mass, particularly marked when Al is involved.
In the line of though of a gas reservoir polluted by matter processed by a
previous generation of stars, out of which the second stellar generation might
form, this correlation may indicate a better ability of more massive
clusters to retain the ejecta.

However, things are not so straightforward, since 
the present day mass of clusters may be different
from the one at the epoch of cluster formation (some 10-12 Gyr ago). 
On the other hand, 
the influence of disruptive processes following the cluster formation does not 
seem
to be significant. In this respect, the paradigm case is provided by the low 
mass globular cluster Pal 5, where \citet{smi02} found evidence of chemical
anomalies similar to those observed in more massive clusters. Yet, Pal 5 has a
mass of only about a few 10$^4$ M$_\odot$, although its tidal tails suggest
that it once was much more massive. 
It is thus possible that the correlation with the mass of the cluster might indeed 
be the record of its past capability to retain and transmute a larger
amount of polluted matter into stars.

The trend for chemical anomalies to increase with increasing ellipticity of 
the cluster is also well
visible in almost all the element ratios, and quite evident for the spread in
the Na-O and Mg-Al anticorrelations, although the uncertain value for
IQR[Mg/Al] in M~5 is somewhat masking the latter trend.
This is not a completely new result, because it is reminiscent of a positive
correlation found to exist \citep{nor87} between the degree of CN enrichment
in RGB stars of a given cluster and the apparent flattening of that system.
The so-called CN-signature was among the first ones to be noticed among the
outcomes of $p-$capture reactions in H-burning: the enhancement of the 
CN band-strength is explained as the conversion of C into N in the CN-cycle.
We must however recall that no dependence on present-day cluster mass was found
for the degree of cyanogen enhancement, which involves the lighter species in 
the $p-$capture chains \citep{smi81}.

At that time, the observed correlation was though to be established via
transfer of angular momentum from individual stars to the system: an higher (on
average) individual angular momentum could result into higher systematic
cluster rotation and consequently into a more marked flattening.
The bottom line was that internal stellar rotation could be responsible for both
the overall shape of the cluster and its degree of chemical inhomogeneity
through internal mixing.
At present, the correlations in Figure~\ref{f12} would be rather hard 
to explain in the scenario where the bulk of anomalies is likely to be
primordial: it is not easy to see why an enhanced degree of pollution should
necessarily be associated to a more flattened cluster.

However, we note that in our sample more massive clusters also show 
a more pronounced ellipticity \citep{har96}: it is possible that the two correlations with
cluster mass and flattening are two expression of the same phenomenon.

While we cannot completely rule out that the observed correlation is simply measuring the
modulation of evolutionary effects in the cluster red giants, we cannot draw
any firmer conclusion before the same IQRs are available for significantly
large samples of unevolved stars, unaffected by any kind of mixing.

Moreover, when looking for possible explanations, we came to notice that 
the shape of clusters, as represented by their
ellipticities, seems to be strongly related to their orbital characteristics. 
This link appears to have been neglected in previous comprehensive reviews 
on various properties of Galactic GCs (for instance in \citealt{djor94}), likely
due to the lack of good sets of absolute proper motions, at that time.
We will discuss such newly found correlations in the next Section.

\subsection{Chemical anomalies and orbital parameters: a true surprise?}

The inference from the previous Section is that the spread of abundances for
elements involved in $p-$capture reactions is more or less congenital in each
globular cluster; the next logical step is thus to look for phenomena that
might interfere with or modify the primordial pattern.
The first idea is thus to look at the parameters that locate a cluster into the
Galaxy.

A growing body of accurate absolute proper motions is presently available for
globular clusters, and this allows to derive reliable orbital parameters for 
these objects. We have taken values from the most recent and extensive work  by
\citet{din99} to explore possible relationships between chemical
inhomogeneities  and the features of the cluster orbits.

Unfortunately there are no proper motions available yet for NGC 2808, hence
this cluster cannot be included in this comparison. For three of the clusters
in Table~\ref{tab6} (M~3, M~5, M~15) \citet{din99} note that systematic errors
may dominate over their formal error estimates and they caution that the orbits
are poorly constrained, although some informations such as the total energy
of the cluster and the shape of the orbit are still useful. Among these
objects, they note that clusters with large total energy (for instance M 5) may
have more  uncertain orbital parameters.

\begin{figure}
\epsscale{1.0}
\plotone{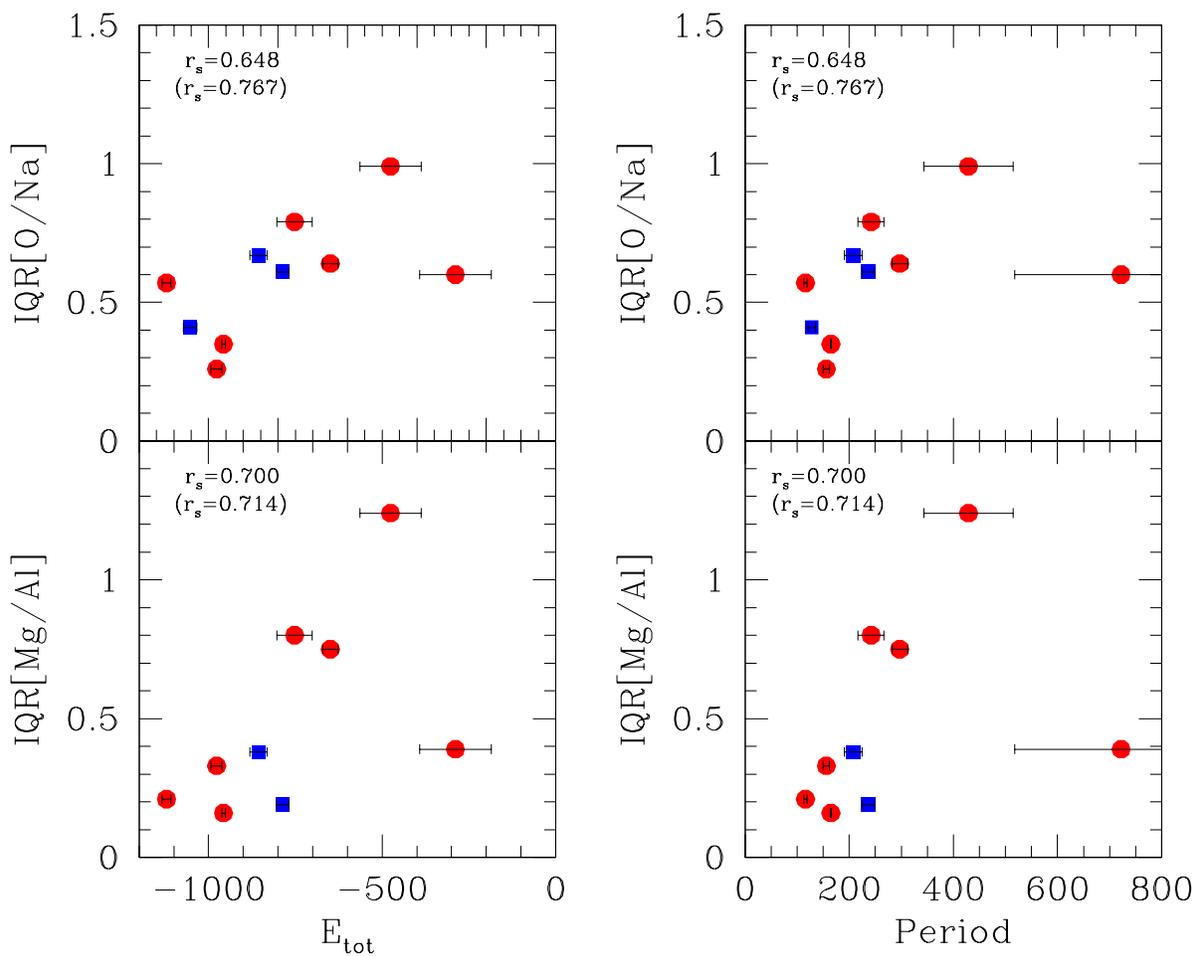}
\caption{IQRs along the Na-O and Mg-Al anticorrelations in globular clusters 
as a function of
the total energy E$_{tot}$ (in units of 10$^2$ km$^2$ s$^{-2}$; left panels) 
and of the period P (in units of 10$^6$ yrs; right panels) of the cluster 
orbit. Blue squares are clusters with less than 20 stars measured. The
Spearman rank coefficient is shown in each panel.\label{f14}}
\end{figure}

In Figure~\ref{f14} we show the correlations we found for the clusters in our
sample (filled circles) between the amount of chemical anomalies in the
Na-O and Mg-Al anticorrelations and orbital parameters.
There is a quite good correlation between IQR[O/Na],
IQR[Mg/Al] and both the total energy $E_{tot}$ and the period P of revolution
around the $z-$axis, in particular when disregarding M 5.
{\it This is the first time, to our knowledge, that such a link between
chemistry of p-capture elements and orbital characteristics is found}.

To test the reliability of this finding, we relaxed somewhat our selection
criteria, and in order to add more clusters we computed IQR[O/Na] and
IQR[Mg/Al] for three other systems having less than $\sim 20$ stars analyzed per
cluster, namely NGC 288, NGC 362 \citep{shet00} and M 10 \citep{kra95}.
The values found for IQR[O/Na] (IQR[Mg/Al]) are +0.61 (0.19), +0.67 (+0.38) and
0.41\footnote{For M 10 no Al abundances were derived by \citet{kra95}.} for NGC
288, NGC 362 and M 10, respectively. When added in Figure~\ref{f14} 
(squares), these three additional clusters nicely follow the trend defined by
our ``main" sample: the largest spread along the Na-O and Mg-Al
anticorrelations is for clusters with the largest period P and $E_{tot}$. In
turn, $E_{tot}$ gives a measure of the major semi axis of the orbit, hence
Figure~\ref{f14} is showing that globular clusters with large-sized
orbits have an enhanced impact of pollution in element from $p-$capture
reactions, with respect to the clusters with small-sized orbits.

The correlation is fairly well defined (again, the statistical significance 
is higher when disregarding M 5): the spread
increases by a factor of $\sim 6$ in [O/Na] and by about one order of magnitude in
[Mg/Al] when orbital parameters increase by a factor of 2.

\begin{figure}
\epsscale{1.0}
\plotone{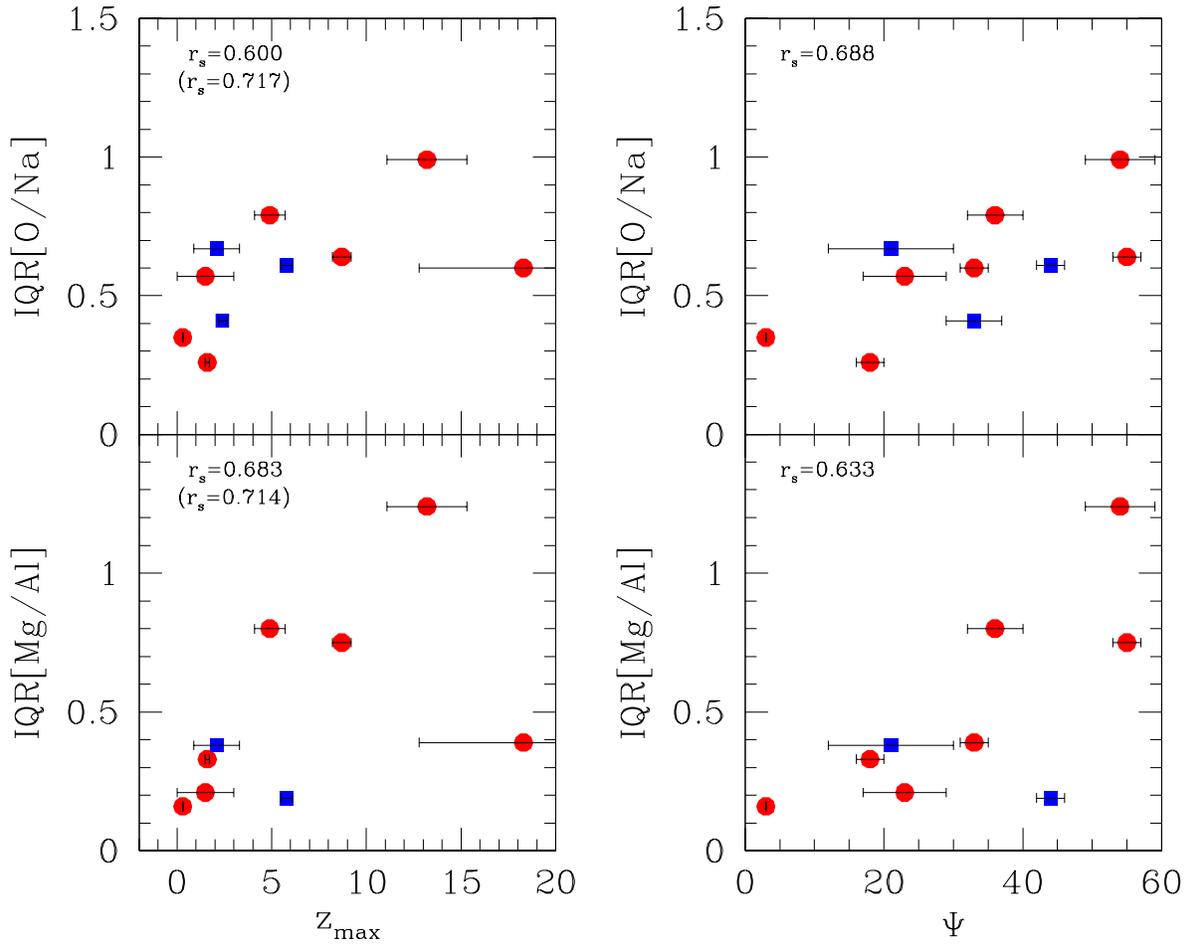}
\caption{IQRs along the Na-O and Mg-Al anticorrelations in globular clusters 
as a function the maximum height above the galactic plane (in kpc; left panels)
and of the orbit inclination (degrees; right panels). Symbols are as in the
previous figure.\label{f15}}
\end{figure}

The same chemical indicators are also plotted in Figure~\ref{f15} as a
function of the maximum height above the plane $z_{max}$ and of the inclination
angle $\Psi$ with respect to the Galactic plane. Again, apart from the
straggler M 5 with its larger uncertainties in derived orbital parameters, all
clusters partake of a quite good correlation between increasing IQRs (in both
Na-O and Mg-Al) as the parameters $z_{max}$ and $\Psi$ increase.

We note that all the subgroups in the \citet{zinn96} classification 
(disk, RHB, BHB and metal-poor clusters) are represented in our sample of 
Table~\ref{tab6} (plus additional clusters): clusters of any metallicity or HB
type lie onto these relations, regardless of their membership to
the disk or halo kinematical subgroups (for instance, both NGC 6752 and M 10
are metal-poor, yet they show a thick disk-like kinematics, according to
\citealt{din99}).

What is the meaning of these up to now unexplored correlations? 
A direct inference from these diagrams is that the largest chemical
inhomogeneities are observed in clusters that spend most of their lifetime
farther away from the Galactic plane, thanks to their large-sized orbits
($E_{tot}$, P) and/or their orbital shapes ($\Psi$, $z_{max}$). Thus, the
simplest way to explain this behavior would seem to postulate that clusters
less affected by the major disturbance due to interactions with the disk are
also less prone to be stripped of a large amount of their gas reservoir.

However, these orbital parameters are determined as averages over a number of
cycles on a 10 Gyr integration time \citep{din99}. On the other hand, stars
presently participating in the Na-O or Mg-Al anticorrelations must have formed
rather early and in a short time in the cluster history: the narrowness of
the main sequence in the color-magnitude diagrams of GCs and 
the old ages derived from the turn-off points strongly support that low mass 
stars in clusters are coeval and old.
Hence, the observed relations with the average parameters should be better
interpreted as a consequence of the proto-clusters being $formed$ away from the
plane and afterward left relatively undisturbed for long intervals of time.
In this respect, Figure~\ref{f14} and Figure~\ref{f15} may be
considered a snapshot of the initial conditions (in both chemistry and
kinematics) at the epoch of the major burst of star formation in globular
clusters.

\begin{figure}
\epsscale{1.0}
\plotone{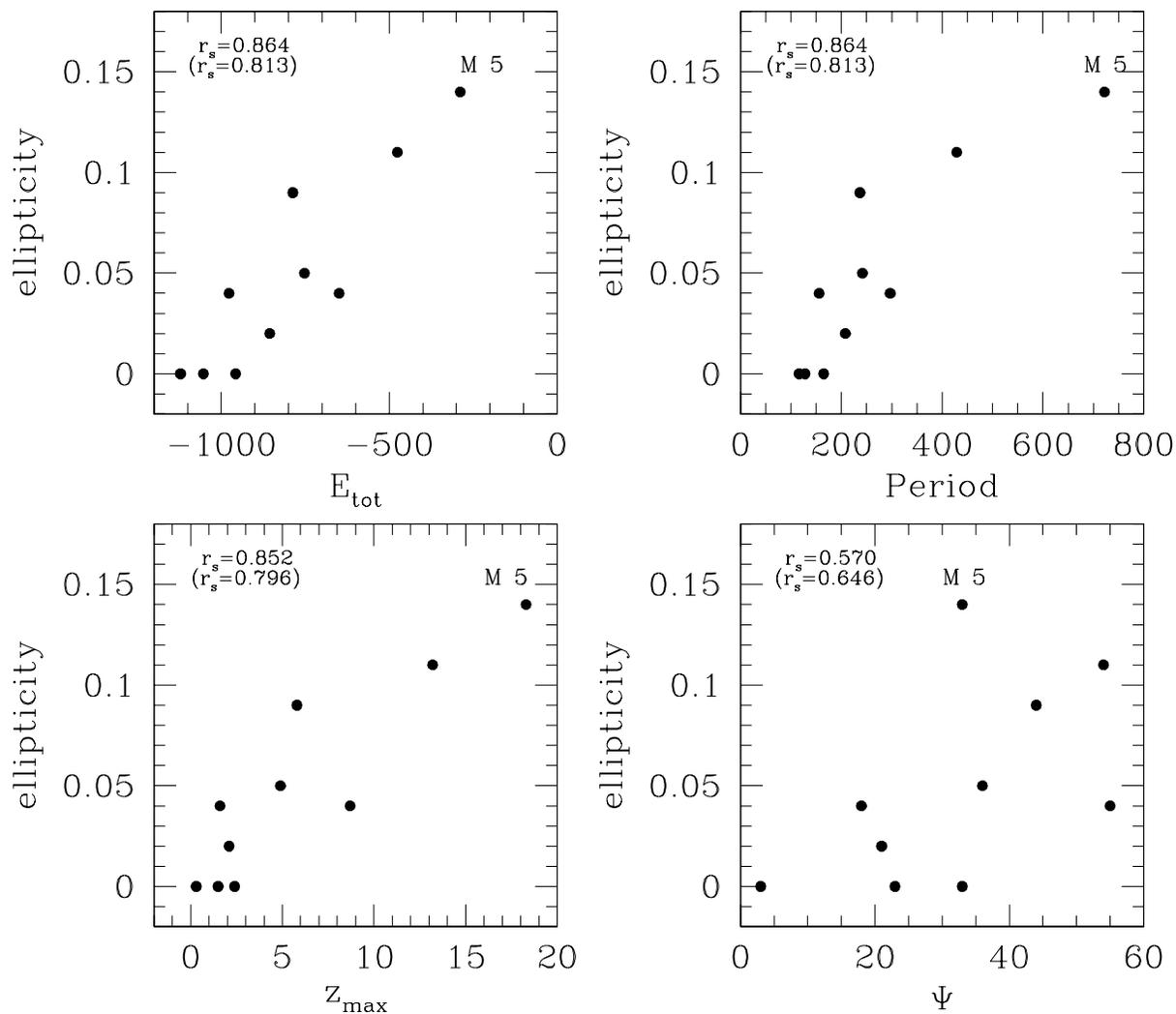}
\caption{Ellipticities of globular clusters in our sample as a function of
orbital parameters $E_{tot}$, period P, 
maximum height above the galactic plane 
and orbit inclination. The globular cluster M 5 (with more uncertain orbital
parameters) is indicated in each panel.\label{f16}}
\end{figure}

This view is supported also by relations we found between structural and orbital
parameters. We plot in Figure~\ref{f16} the values of cluster ellipticities 
from \citet{har96} as a function of orbital parameters $E_{tot}$, P,
$z_{max}$ and inclination angle $\Psi$. In these plots we highlight M~5 as the
cluster with larger uncertainties in derived orbital parameters; instead, the
three additional clusters are indicated with the same symbols as the clusters
in our main sample.

There is a quite good correlation between the flattening of clusters and,
again, quantities that, in a sense, are measuring how much time the system may
spend far away from the major disturbance of the Galactic disk. A possible
interpretation is that near the Galactic plane the effects of
relaxation which acts to cancel out the anisotropy of orbits in a given
cluster are more important. The net effect would then be that 
clusters spending most of their
lifetime far from the disk would also have a greater chance to maintain a
certain degree of flattening.
At the same time, these objects on highly energetic orbits (and/or with large 
inclination about the plane) are also able to better save the memory of the
initial conditions in which their presently observed low mass stars formed.

\section{Summary and conclusions}

We analyzed the chemical abundances of 19 red giant stars in the globular
cluster NGC 2808. Abundance ratios for Al, $\alpha-$process and Fe-group
elements have been derived for stars from the tip of RGB down below the
RGB-bump. We found no trend of abundance ratios as a function of effective
temperatures.

Elements involved in the NeNa, MgAl cycles of $p-$capture reactions in high
temperature H-burning show large star-to-star variations in this cluster.
There is evidence that synthesis of Na and Al in these reactions was large
enough to leave some stars in NGC 2808 heavily polluted by the outcome of this
burning. This cluster hosts at least some stars super O-poor as their
counterparts in M 13, the paradigm clusters for large chemical inhomogeneities.
These stars are also Mg-depleted and Na,Al-enhanced, just the kind of pattern
expected.

We found that the correlation of Na and Al abundances seems to be different in
slope among clusters of similar metallicity such as NGC 2808, M 4 and M 5.

We defined the interquartile range of the [X/Fe] distribution as a good
quantitative measure of the degree of inhomogeneities in a cluster.
Using a set of clusters with abundance analysis from high resolution
spectroscopy in a fairly large number of star per cluster, we explored possible
relations with global cluster parameters. None is found with cluster
metallicity, age or HB morphological type.

Rather good correlations are found with both the present day mass of clusters
and with the cluster flattening as measured by ellipticity. We note that these
two quantities are correlated with each other in our cluster sample.

The most striking relationships are found, for the first time to our knowledge,
between the amount of spread along the Na-O and Mg-Al anticorrelations and
orbital parameters of the clusters: the larger the orbit sizes and revolution
periods, the larger the amount of inhomogeneities. Larger spreads in [O/Na]
and [Mg/Al] distributions are clearly found also for clusters with large 
maximum heights
above the Galactic plane and/or with larger inclination angles of the orbit
with respect to the plane.

We put forward the working hypothesis that these correlations are a reflex of
the initial conditions at the time the presently observed cluster stars formed
out of the proto cluster gas, likely polluted by the ejecta of a prior generation
of more massive stars. In GCs later spending most of their life far away
from the major disturbance of close interactions with the Galactic disk there
is a greater chance that most of the gas reservoir was retained to form stars
showing large signature of pollution. 
This view is also well supported by the observed correlations found between
cluster ellipticity and orbital parameters: we regard these tight relations as
evidence of a reduced action of relaxation processes in erasing the anisotropy
of stellar orbits in clusters standing far from the disk for large periods of
their lifetime.

The emerging scenario is that a complex interaction of initial
conditions, dynamical and chemical evolution might be required to explain all
the features and relations found for the global parameters in globular
clusters. 

\acknowledgments
This publication makes use of data products from the Two Micron All Sky Survey,
which is a joint project of the University of Massachusetts and the Infrared
Processing and Analysis Center/California Institute of Technology, funded by
the National Aeronautics and Space Administration and the National Science
Foundation. 
I wish to warmly thank Angela Bragaglia for enlightening comments and a careful
reading of the manuscript, Micol Bolzonella, Barbara Lanzoni, Luca Ciotti 
for useful and kind discussions, and Lucia Ballo and her Don Quijote 
for constant encouragement and for being herself. I also thank the anonymous 
referee whose suggestions and comments helped to improve this paper.
This work was partially funded by Cofin 2003-029437 and Cofin 2004-025729 
from the Italian MIUR.

\clearpage

\begin{deluxetable}{rrcccrccrcc}
\tablecaption{Adopted atmopheric parameters and Fe abundances.\label{tab1}}
\tablehead{\colhead{Star ID}&
\colhead{$S/N$}&
\colhead{T$_{\rm eff}$}&
\colhead{$\log g$}&
\colhead{$v_t$}&
\colhead{n}&
\colhead{[Fe/H]I}&
\colhead{$\sigma$}&
\colhead{n}&
\colhead{[Fe/H]II}&
\colhead{$\sigma$}\\
\colhead{}&
\colhead{}&
\colhead{K}&
\colhead{dex}&
\colhead{km s$^{-1}$}&
\colhead{}&
\colhead{dex}&
\colhead{dex}&
\colhead{}&
\colhead{dex}&
\colhead{dex}\\
}
\startdata
10201 &   45 & 4717 &2.02 &1.20 & 51 &$-$1.06 &0.13&  8&$-$1.21& 0.05 \\ 
13983 &   40 & 4826 &2.17 &0.60 & 33 &$-$1.08 &0.08&  5&$-$1.09& 0.13 \\ 
32685 &   30 & 4788 &2.03 &0.83 & 34 &$-$1.15 &0.09&  5&$-$1.41& 0.07 \\ 
34013\tablenotemark{a} &   20 & 5110 &2.51 &0.80 & 29 &$-$0.89 &0.21&  3&$-$1.84& 0.12 \\ 
37872 &  120 & 4015 &0.71 &1.68 & 96 &$-$1.10 &0.11& 14&$-$1.06& 0.08 \\ 
42886 &   30 & 4791 &2.14 &0.85 & 47 &$-$1.16 &0.14&  4&$-$1.38& 0.10 \\ 
43217 &   30 & 4916 &2.41 &0.80 & 33 &$-$1.00 &0.12&  7&$-$1.17& 0.13 \\ 
46099 &   80 & 4032 &0.76 &1.72 & 60 &$-$1.18 &0.12& 10&$-$1.19& 0.12 \\ 
46422 &  100 & 3943 &0.52 &1.85 & 95 &$-$1.17 &0.12& 14&$-$1.08& 0.09 \\ 
46580 &   65 & 4051 &0.74 &1.68 & 89 &$-$1.15 &0.11& 15&$-$1.03& 0.13 \\ 
47606 &  110 & 3839 &0.44 &1.66 & 85 &$-$1.12 &0.14& 14&$-$1.14& 0.16 \\ 
48609 &  110 & 3846 &0.44 &1.78 & 58 &$-$1.22 &0.12&  8&$-$1.11& 0.13 \\ 
48889 &   85 & 3943 &0.52 &1.80 & 99 &$-$1.15 &0.15& 13&$-$1.22& 0.12 \\ 
50119 &   70 & 4166 &0.93 &1.73 &119 &$-$1.08 &0.15& 18&$-$1.21& 0.13 \\ 
50761 &  120 & 3756 &0.31 &1.75 & 81 &$-$1.22 &0.12& 10&$-$0.81& 0.07 \\ 
51454 &  120 & 3893 &0.51 &1.65 & 90 &$-$1.26 &0.11& 14&$-$1.11& 0.13 \\ 
51499 &   85 & 3960 &0.57 &1.70 &104 &$-$1.25 &0.10& 18&$-$1.24& 0.12 \\ 
51983 &   95 & 3855 &0.47 &1.77 & 71 &$-$1.15 &0.11& 13&$-$1.16& 0.12 \\ 
53390 &   60 & 4426 &1.43 &1.30 &106 &$-$1.12 &0.10& 19&$-$1.14& 0.13 \\ 
56032 &   70 & 4045 &0.87 &1.70 &103 &$-$1.10 &0.10& 17&$-$0.99& 0.12 \\ 
\enddata
\tablenotetext{a}{For this star measurements of $EW$s were scarcely reliable,
hence this star is disregarded from further discussion and is reported here
only for the sake of completeness.}
\end{deluxetable}

\clearpage

\begin{deluxetable}{lrclrclrclrclrclrclr}
\tabletypesize{\scriptsize}
\rotate
\tablecaption{Abundances of Al and $\alpha$-elements in stars of NGC 2808.\label{tab2}}
\tablewidth{0pt}
\tablehead{
\colhead{Star} & \colhead{n} &\colhead{[Al/Fe]} & \colhead{$\sigma$} & 
\colhead{n} &\colhead{[Mg/Fe]} & \colhead{$\sigma$} & 
\colhead{n} &\colhead{[Si/Fe]} & \colhead{$\sigma$} &
\colhead{n} &\colhead{[Ca/Fe]} & \colhead{$\sigma$} & 
\colhead{n} &\colhead{[Ti/Fe]I}& \colhead{$\sigma$} & 
\colhead{n} &\colhead{[Ti/Fe]II}&\colhead{$\sigma$}
}
\startdata
10201 & 2 &  +0.92 & 0.12    & 3 &  +0.29 & 0.19  & 5 & +0.39 & 0.11 & 15 & +0.25 & 0.20 & 16 & +0.20 &0.21 & 2 & +0.26 & 0.01   \\
13983 & 2 &  +0.94 & 0.02    & 2 &  +0.24 & 0.09  & 5 & +0.39 & 0.08 & 16 & +0.28 & 0.15 &  8 & +0.32 &0.19 & 0 &\nodata &\nodata\\
32685 & 2 &  +0.86 & 0.09    & 1 &  +0.47 &\nodata& 5 & +0.36 & 0.08 &  9 & +0.36 & 0.17 &  9 & +0.31 &0.20 & 3 & +0.19 & 0.13   \\
37872 & 2 &  +1.03 & 0.04    & 4 &  +0.21 & 0.26  & 7 & +0.39 & 0.11 & 17 & +0.33 & 0.12 & 12 & +0.24 &0.08 & 7 & +0.12 & 0.16   \\
42886 & 2 &$-$0.07 & 0.09    & 2 &  +0.40 & 0.24  & 2 & +0.32 & 0.33 & 17 & +0.31 & 0.20 & 11 & +0.39 &0.23 & 3 & +0.23 & 0.19   \\
43217 & 0 & \nodata& \nodata & 2 &  +0.44 & 0.15  & 1 & +0.12&\nodata& 15 & +0.32 & 0.22 & 10 & +0.35 &0.21 & 2 & +0.30 & 0.11   \\
46099 & 2 &  +0.16 & 0.08    & 4 &  +0.40 & 0.28  & 5 & +0.29 & 0.08 & 12 & +0.33 & 0.12 & 18 & +0.36 &0.16 & 7 & +0.23 & 0.17   \\
46422 & 2 &  +0.19 & 0.00    & 4 &  +0.48 & 0.09  & 6 & +0.35 & 0.10 & 14 & +0.33 & 0.19 & 17 & +0.30 &0.18 & 6 & +0.18 & 0.08   \\
46580 & 2 &  +0.32 & 0.01    & 3 &  +0.33 & 0.12  & 6 & +0.27 & 0.07 & 16 & +0.32 & 0.10 & 20 & +0.36 &0.18 & 5 & +0.15 & 0.13   \\
47606 & 2 &$-$0.01 & 0.08    & 3 &  +0.19 & 0.05  & 7 & +0.42 & 0.09 & 13 & +0.31 & 0.12 & 16 & +0.28 &0.13 & 5 & +0.14 & 0.09   \\
48609 & 2 &  +0.14 & 0.06    & 4 &  +0.51 & 0.26  & 7 & +0.42 & 0.18 & 11 & +0.31 & 0.15 & 14 & +0.36 &0.11 & 6 & +0.27 & 0.07   \\
48889 & 2 &  +1.02 & 0.19    & 3 &  +0.32 & 0.19  & 6 & +0.51 & 0.15 & 18 & +0.42 & 0.16 & 18 & +0.41 &0.14 & 5 & +0.25 & 0.15   \\
50119 & 2 &  +1.42 & 0.06    & 3 &$-$0.23 & 0.12  & 8 & +0.40 & 0.12 & 17 & +0.38 & 0.17 & 25 & +0.20 &0.14 & 5 & +0.26 & 0.19   \\
50761 & 2 &  +0.29 & 0.04    & 4 &  +0.33 & 0.22  & 7 & +0.40 & 0.06 & 14 & +0.18 & 0.14 & 16 & +0.28 &0.23 & 5 & +0.19 & 0.19   \\
51454 & 2 &  +0.11 & 0.05    & 3 &  +0.40 & 0.07  & 7 & +0.41 & 0.11 & 18 & +0.38 & 0.14 & 16 & +0.37 &0.14 & 7 & +0.20 & 0.22   \\
51499 & 2 &  +0.05 & 0.05    & 3 &  +0.46 & 0.20  & 6 & +0.37 & 0.11 & 15 & +0.31 & 0.10 & 18 & +0.36 &0.19 & 4 & +0.27 & 0.07   \\
51983 & 2 &  +0.91 & 0.19    & 4 &  +0.25 & 0.27  & 8 & +0.40 & 0.18 & 16 & +0.42 & 0.13 & 18 & +0.37 &0.18 & 4 & +0.12 & 0.06   \\
53390 & 2 &$-$0.25 & 0.12    & 4 &  +0.33 & 0.07  & 8 & +0.32 & 0.07 & 17 & +0.31 & 0.13 & 24 & +0.24 &0.17 & 8 & +0.17 & 0.19   \\
56032 & 2 &  +0.07 & 0.05    & 4 &  +0.39 & 0.07  & 8 & +0.41 & 0.18 & 15 & +0.30 & 0.14 & 17 & +0.19 &0.08 & 6 & +0.15 & 0.16   \\
\enddata
\end{deluxetable}

\clearpage
\begin{deluxetable}{lrclrclrclrclrclrclr}
\tabletypesize{\scriptsize}
\rotate
\tablecaption{Abundances of Fe-group elements in stars of NGC 2808.\label{tab3}}
\tablewidth{0pt}
\tablehead{
\colhead{Star} & \colhead{n} &\colhead{[Sc/Fe]II} & \colhead{$\sigma$} & 
\colhead{n} &\colhead{[V/Fe]} & \colhead{$\sigma$} & 
\colhead{n} &\colhead{[Cr/Fe]I} & \colhead{$\sigma$} &
\colhead{n} &\colhead{[Cr/Fe]II} & \colhead{$\sigma$} & 
\colhead{n} &\colhead{[Mn/Fe]}& \colhead{$\sigma$} & 
\colhead{n} &\colhead{[Ni/Fe]}&\colhead{$\sigma$}
}
\startdata
10201 & 6 &  +0.12 & 0.11   & 6 &$-$0.13 & 0.06 &  8 &$-$0.11 & 0.23  &  4 &  +0.10 & 0.19 &   5 &$-$0.45 & 0.13 &   13 &$-$0.11 & 0.18\\
13983 & 1 &$-$0.08 &\nodata & 4 &$-$0.13 & 0.11 &  4 &  +0.05 & 0.09  &  2 &$-$0.00 & 0.21 &   6 &$-$0.40 & 0.14 &    5 &$-$0.08 & 0.14\\
32685 & 5 &  +0.16 & 0.18   & 3 &  +0.01 & 0.14 &  8 &  +0.08 & 0.17  &  0 &\nodata&\nodata&   5 &$-$0.30 & 0.12 &   11 &$-$0.07 & 0.21\\
37872 & 6 &  +0.01 & 0.10   & 5 &$-$0.01 & 0.17 & 13 &  +0.04 & 0.18  &  5 &  +0.14 & 0.18 &   4 &$-$0.23 & 0.15 &   31 &$-$0.07 & 0.10\\
42886 & 5 &  +0.02 & 0.07   & 4 &  +0.15 & 0.21 &  4 &  +0.12 & 0.20  &  2 &  +0.15 & 0.06 &   6 &$-$0.33 & 0.06 &    8 &$-$0.04 & 0.11\\
43217 & 5 &  +0.02 & 0.22   & 5 &$-$0.11 & 0.05 &  5 &  +0.12 & 0.11  &  1 &  +0.19 &\nodata&  6 &$-$0.51 & 0.13 &    7 &$-$0.10 & 0.17\\
46099 & 7 &  +0.10 & 0.14   & 5 &  +0.07 & 0.14 & 10 &$-$0.02 & 0.17  &  2 &  +0.03 & 0.29 &   4 &$-$0.11 & 0.06 &   25 &$-$0.09 & 0.08\\
46422 & 7 &$-$0.17 & 0.09   & 6 &  +0.16 & 0.22 & 12 &  +0.06 & 0.15  &  4 &  +0.06 & 0.11 &   3 &$-$0.21 & 0.12 &   32 &$-$0.06 & 0.14\\
46580 & 7 &$-$0.10 & 0.07   & 7 &$-$0.04 & 0.17 & 14 &  +0.08 & 0.18  &  4 &  +0.09 & 0.08 &   4 &$-$0.28 & 0.04 &   32 &$-$0.03 & 0.12\\
47606 & 7 &$-$0.11 & 0.10   & 6 &  +0.18 & 0.19 & 10 &  +0.10 & 0.14  &  2 &  +0.19 & 0.21 &   3 &$-$0.13 & 0.31 &   25 &$-$0.09 & 0.14\\
48609 & 7 &$-$0.04 & 0.10   & 4 &  +0.15 & 0.18 &  9 &$-$0.02 & 0.18  &  2 &$-$0.01 & 0.08 &   3 &$-$0.23 & 0.15 &   25 &$-$0.03 & 0.15\\
48889 & 7 &  +0.06 & 0.08   & 5 &  +0.20 & 0.15 & 10 &  +0.05 & 0.09  &  3 &  +0.13 & 0.22 &   3 &$-$0.19 & 0.07 &   32 &$-$0.14 & 0.14\\
50119 & 6 &  +0.23 & 0.12   & 7 &  +0.13 & 0.17 & 12 &$-$0.05 & 0.07  &  3 &  +0.21 & 0.16 &   4 &$-$0.23 & 0.07 &   35 &$-$0.10 & 0.16\\
50761 & 6 &$-$0.60 & 0.13   & 5 &$-$0.02 & 0.16 &  8 &$-$0.00 & 0.17  &  2 &  +0.04 & 0.19 &   4 &$-$0.28 & 0.21 &   30 &$-$0.05 & 0.17\\
51454 & 6 &$-$0.17 & 0.07   & 4 &$-$0.07 & 0.10 &  8 &  +0.13 & 0.08  &  3 &  +0.17 & 0.22 &   4 &$-$0.29 & 0.24 &   38 &  +0.01 & 0.16\\
51499 & 7 &  +0.08 & 0.12   & 7 &  +0.05 & 0.16 & 14 &$-$0.03 & 0.09  &  5 &  +0.19 & 0.21 &   3 &$-$0.14 & 0.04 &   30 &$-$0.09 & 0.08\\
51983 & 7 &$-$0.00 & 0.13   & 6 &  +0.19 & 0.07 & 13 &$-$0.06 & 0.10  &  2 &$-$0.03 & 0.03 &   3 &$-$0.20 & 0.24 &   24 &$-$0.13 & 0.13\\
53390 & 6 &$-$0.01 & 0.06   &10 &  +0.01 & 0.15 & 15 &  +0.03 & 0.15  &  5 &  +0.02 & 0.14 &   4 &$-$0.34 & 0.11 &   33 &$-$0.02 & 0.12\\
56032 & 7 &$-$0.18 & 0.07   & 7 &  +0.04 & 0.13 & 17 &$-$0.05 & 0.17  &  3 &$-$0.10 & 0.18 &   4 &$-$0.40 & 0.16 &   36 &$-$0.04 & 0.17\\
\enddata
\end{deluxetable}

\clearpage
{\tiny
\begin{table}
\begin{center}
\caption{Sensitivities of abundance ratios to errors in the atmospheric 
parameters and in the equivalent widths}\label{tab4}
\begin{tabular}{lccccrcc}
\tableline
Ratio    & $\Delta T_{eff}$ & $\Delta$ $\log g$ & $\Delta$ [A/H] & $\Delta v_t$ &$<N>$& $\Delta$ EW & tot.\\
         & (+70 K)    & (+0.1 dex)      & (+0.1 dex)      & (+0.1 km/s) & &                 & (dex)  \\
\tableline
\\
& \multicolumn{7}{c}{Star 48889: $V=13.34$, T$_{\rm eff}$/$\log g$/$v_t$=3943/0.52/1.80} \\
\cline{2-8} \\
$[$Mg/Fe$]$I &   +0.001& $-$0.017 &$-$0.005 &  +0.008&  4&   +0.058& 0.061 \\
$[$Al/Fe$]$I &   +0.044& $-$0.012 &$-$0.019 &  +0.009&  2&   +0.082& 0.096 \\
$[$Si/Fe$]$I & $-$0.081&   +0.012 &  +0.005 &  +0.024&  7&   +0.044& 0.096 \\
$[$Ca/Fe$]$I &   +0.069& $-$0.021 &$-$0.022 &$-$0.031& 15&   +0.030& 0.087 \\
$[$Sc/Fe$]$II&   +0.085& $-$0.013 &$-$0.005 &$-$0.022&  7&   +0.044& 0.099 \\
$[$Ti/Fe$]$I &   +0.116& $-$0.014 &$-$0.022 &$-$0.022& 18&   +0.027& 0.123 \\
$[$Ti/Fe$]$II&   +0.064& $-$0.019 &$-$0.010 &$-$0.019&  6&   +0.047& 0.085 \\
$[$V/Fe$]$I  &   +0.127& $-$0.008 &$-$0.017 &$-$0.044&  6&   +0.047& 0.143 \\
$[$Cr/Fe$]$I &   +0.078& $-$0.020 &$-$0.020 &  +0.000& 12&   +0.033& 0.090 \\
$[$Cr/Fe$]$II&   +0.023& $-$0.016 &$-$0.017 &  +0.009&  3&   +0.067& 0.075 \\
$[$Mn/Fe$]$I &   +0.036& $-$0.006 &$-$0.010 &$-$0.014&  4&   +0.058& 0.071 \\
$[$Fe/H$]$I  &   +0.025&   +0.010 &  +0.012 &$-$0.041& 90&   +0.012& 0.052 \\
$[$Fe/H$]$II & $-$0.104&   +0.049 &  +0.033 &$-$0.022& 14&   +0.031& 0.125 \\
$[$Ni/Fe$]$I & $-$0.021&   +0.008 &  +0.004 &  +0.008& 31&   +0.021& 0.032 \\

\\
\tableline
\\
& \multicolumn{7}{c}{Star 43217: $V=16.44$, T$_{\rm eff}$/$\log g$/$v_t$=4916/2.41/0.80}\\
\cline{2-8} \\
$[$Mg/Fe$]$I & $-$0.013 &$-$0.026 &  +0.011 &  +0.027&  3&   +0.067& 0.079\\
$[$Al/Fe$]$I & $-$0.039 &  +0.006 &$-$0.001 &  +0.031&  2&   +0.082& 0.096\\
$[$Si/Fe$]$I & $-$0.075 &  +0.020 &  +0.008 &  +0.035&  3&   +0.067& 0.109\\
$[$Ca/Fe$]$I & $-$0.014 &$-$0.010 &$-$0.003 &  +0.007& 14&   +0.031& 0.036\\
$[$Sc/Fe$]$II&   +0.027 &  +0.000 &  +0.000 &  +0.009&  4&   +0.058& 0.064\\
$[$Ti/Fe$]$I &   +0.018 &  +0.003 &$-$0.007 &  +0.008& 10&   +0.037& 0.042\\
$[$Ti/Fe$]$II&   +0.025 &$-$0.002 &$-$0.004 &$-$0.035&  2&   +0.082& 0.093\\
$[$V/Fe$]$I  &   +0.015 &  +0.005 &$-$0.002 &  +0.030&  4&   +0.058& 0.067\\
$[$Cr/Fe$]$I &   +0.019 &$-$0.002 &$-$0.006 &$-$0.010&  6&   +0.047& 0.052\\
$[$Cr/Fe$]$II& $-$0.001 &  +0.000 &$-$0.007 &  +0.011&  2&   +0.082& 0.083\\
$[$Mn/Fe$]$I &   +0.003 &  +0.008 &$-$0.002 &  +0.023&  5&   +0.052& 0.057\\
$[$Fe/H$]$I  &   +0.090 &$-$0.011 &$-$0.005 &$-$0.039& 38&   +0.019& 0.100\\
$[$Fe/H$]$II & $-$0.027 &  +0.040 &  +0.029 &$-$0.030&  5&   +0.052& 0.082\\
$[$Ni/Fe$]$I & $-$0.011 &  +0.013 &  +0.003 &  +0.011&  8&   +0.041& 0.046\\
\tableline
\end{tabular}
\end{center}
\end{table}
}

\clearpage
\begin{deluxetable}{lrclc}
\tablecaption{Mean abundance ratios for stars along the RGB in 
NGC~2808\label{tab5}}
\tablewidth{0pt}
\tablehead{
\colhead{Ratio} &
\colhead{N$_{\rm star}$} &
\colhead{Mean} &
\colhead{$\sigma_{obs}$} &
\colhead{$\sigma_{exp}$}\\
}
\startdata
$[$O/Fe$]$\tablenotemark{a}    &19&   +0.01  & 0.37   & 0.14 \\
$[$Na/Fe$]$\tablenotemark{a}   &19&   +0.22  & 0.25   & 0.09 \\
$[$Mg/Fe$]$   &19&   +0.32  & 0.16   & 0.07 \\
$[$Al/Fe$]$   &18&   +0.45  & 0.49   & 0.10 \\
$[$Si/Fe$]$  &19&   +0.37  & 0.08   & 0.10 \\
$[$Ca/Fe$]$  &19&   +0.32  & 0.06   & 0.06 \\
$[$Sc/Fe$]$II &19& $-$0.04  & 0.18   & 0.08 \\
$[$Ti/Fe$]$I  &19&   +0.31  & 0.07   & 0.08 \\
$[$Ti/Fe$]$II &18&   +0.20  & 0.06   & 0.09 \\
$[$V/Fe$]$   &19&   +0.04  & 0.11   & 0.11 \\
$[$Cr/Fe$]$I  &19&   +0.03  & 0.07   & 0.07 \\
$[$Cr/Fe$]$II &18&   +0.09  & 0.09   & 0.08 \\
$[$Mn/Fe$]$  &19& $-$0.27  & 0.10   & 0.06 \\
$[$Fe/H$]$I   &19& $-$1.14  & 0.07   & 0.08 \\
$[$Fe/H$]$II  &19& $-$1.14  & 0.13   & 0.10 \\
$[$Ni/Fe$]$  &19& $-$0.07  & 0.04   & 0.04 \\
\enddata
\tablenotetext{a}{Values used in this average are from \citet{car04nao28}; Na abundances do include corrections for departures from LTE according to
\citet{gra99}} 
\end{deluxetable}

\clearpage
\thispagestyle{empty}
\begin{deluxetable}{lcccccccccccccr}
\tabletypesize{\scriptsize}
\rotate
\tablewidth{0pt}
\tablecaption{Median values and Interquartile Range (IQR) for 8 
globular clusters.\label{tab6}}
\tablehead{
\colhead{Cluster} & \colhead{[Fe/H]\tablenotemark{a}} &
\colhead{median} & \colhead{IQR} &
\colhead{median} & \colhead{IQR} &
\colhead{median} & \colhead{IQR} &
\colhead{median} & \colhead{IQR} &
\colhead{median} & \colhead{IQR} &
\colhead{median} & \colhead{IQR} &
\colhead{ref}\\
\colhead{} & \colhead{} &
\colhead{[O/Fe]} & \colhead{[O/Fe]} &
\colhead{[Na/Fe]} & \colhead{[Na/Fe]} &
\colhead{[Mg/Fe]} & \colhead{[Mg/Fe]} &
\colhead{[Al/Fe]} & \colhead{[Al/Fe]} &
\colhead{[O/Na]} & \colhead{[O/Na]} &
\colhead{[Mg/Al]} & \colhead{[Mg/Al]} &
\colhead{}}
\startdata
NGC 2808     & $-$1.14 &  0.13 & 0.49 &0.17 & 0.37 & 0.33 & 0.19 & 0.29 & 0.85 &$-$0.08 & 0.83 &   +0.20 & 0.98 & 1 \\
NGC 5272 M3  & $-$1.34 &  0.20 & 0.26 &0.28 & 0.34 & 0.39 & 0.22 & 0.86 & 0.64 &$-$0.08 & 0.64 & $-$0.49 & 0.75 & 2 \\
NGC 5904 M5  & $-$1.11 &  0.13 & 0.36 &0.29 & 0.36 & 0.53 & 0.10 & 0.72 & 0.29 &$-$0.18 & 0.60 & $-$0.40\tablenotemark{b} & 0.39\tablenotemark{b} & 3 \\
NGC 6121 M4  & $-$1.19 &  0.38 & 0.20 &0.40 & 0.41 & 0.62 & 0.11 & 0.89 & 0.22 &  +0.01 & 0.57 & $-$0.31 & 0.21 & 4 \\
NGC 6205 M13 & $-$1.39 &  0.02 & 0.46 &0.61 & 0.46 & 0.31 & 0.24 & 1.17 & 1.03 &$-$0.65 & 0.99 & $-$0.57 & 1.24 & 2 \\
NGC 6752     & $-$1.42 &  0.25 & 0.16 &0.44 & 0.18 & 0.67 & 0.03 & 1.04 & 0.28 &$-$0.17 & 0.26 & $-$0.36 & 0.33 & 5 \\
NGC 6838 M71 & $-$0.70 &  0.36 & 0.22 &0.42 & 0.16 & 0.59 & 0.11 & 0.59 & 0.11 &$-$0.07 & 0.35 &   +0.05 & 0.16 & 6 \\
NGC 7078 M15 & $-$2.12 &  0.26 & 0.32 &0.40 & 0.48 & 0.59 & 0.24 & 1.14 & 0.60 &$-$0.10 & 0.79 & $-$0.45 & 0.80 & 7 \\
\enddata
\tablenotetext{a}{Metallicity is either from \citet{car04nao28} or
from \citet{cg97}.}
\tablenotetext{b}{These values are uncertain due to the small number of stars
with both Al and Mg abundances available for M 5.}


\tablerefs{
(1) this study; (2) Sneden et al. (2004); (3) Ivans et al. (2001);
(4) Ivans et al. (1999); (5) Yong et al. (2003): (6) Ramirez \& Cohen (2002);
(7) Sneden et al. (1997)}
\end{deluxetable}

\clearpage



\clearpage

\clearpage

\clearpage

\clearpage

\clearpage

\clearpage

\clearpage

\clearpage

\clearpage

\clearpage

\clearpage

\clearpage

\clearpage

\clearpage

\clearpage

\end{document}